\documentclass[twocolumn,prb,aps,footinbib,amsmath,amssymb,superscriptaddress]{revtex4-1}
\usepackage{graphicx}
\usepackage{dcolumn}
\usepackage{color}
\usepackage{bm}
\usepackage{amsmath,amssymb,amsfonts}
\usepackage{svg}
\usepackage{listings}
\usepackage[caption=false]{subfig}

\newcommand{\lolo}[1] {{\color{magenta}{#1}}\color{black}\normalsize}

\newlength{\subcolumnwidth}

\newcommand{\nextsubcolumn}[1][]{%
  \cr\noalign{\hfill}
  \if\relax\detokenize{#1}\relax\else\hsize=#1\setlength{\subcolumnwidth}{\hsize}\fi
}

\newcommand{\Nc}{NiCp$_2$~}
\newcommand{\Nd}{NiCp$_2$}
\begin{document}

\title{One dimensional chains of nickelocene fragments on  Au(111)}
\author{Divya Jyoti}
\affiliation{Centro de F{\'{i}}sica de Materiales
        CFM/MPC (CSIC-UPV/EHU),  20018 Donostia-San Sebasti\'an, Spain}
\affiliation{Donostia International Physics Center (DIPC),  20018 Donostia-San Sebasti\'an, Spain} 

\author{Alex F\'etida}
\affiliation{{Universit\'{e} de Strasbourg, CNRS, IPCMS, UMR 7504, F-67000 Strasbourg, France}}       
\author{Laurent Limot}
\affiliation{{Universit\'{e} de Strasbourg, CNRS, IPCMS, UMR 7504, F-67000 Strasbourg, France}}
\author{Roberto Robles}
\affiliation{Centro de F{\'{i}}sica de Materiales
        CFM/MPC (CSIC-UPV/EHU),  20018 Donostia-San Sebasti\'an, Spain}
\author{Nicol{\'a}s Lorente}
\affiliation{Centro de F{\'{i}}sica de Materiales
        CFM/MPC (CSIC-UPV/EHU),  20018 Donostia-San Sebasti\'an, Spain}
\affiliation{Donostia International Physics Center (DIPC),  20018 Donostia-San Sebasti\'an, Spain}
\author{Deung-Jang Choi}
\email{djchoi@dipc.org}
\affiliation{Centro de F{\'{i}}sica de Materiales
        CFM/MPC (CSIC-UPV/EHU),  20018 Donostia-San Sebasti\'an, Spain}
\affiliation{Donostia International Physics Center (DIPC),  20018 Donostia-San
 Sebasti\'an, Spain}
 \affiliation{Ikerbasque, Basque Foundation for Science, 48013 Bilbao, Spain}

\begin{abstract}
We investigate the temperature-dependent deposition of nickelocene (\Nd) molecules on a single crystal Au(111) substrate, revealing distinct adsorption behaviors and structural formations. At low temperatures (4.2 K), individual \Nc molecules adsorb on the herringbone elbows and step edges, forming ordered patterns as molecular coverage increases. However, at 77 K, the molecules dissociate, yielding two main fragments: NiCp fragments that are Ni atoms capped by cyclopentadienyl (Cp) rings, which preferentially adsorb at FCC hollow sites, and Cp radical fragments exhibiting strong substrate interactions. NiCp  fragments self-assemble into one-dimensional (1-D) chains along the $\langle 1 1 \bar{2} \rangle$ directions, displaying higher protrusion in STM images. The strain and steric hindrance from the Cp protons induce chiral patterns within the chains, which are well-reproduced by our DFT simulations. In contrast, the Cp fragments maintain distances due to short-range repulsive forces and exhibit low diffusion barriers. Interestingly, the fragments are non-magnetic, as confirmed by both STM measurements and DFT calculations, in contrast to the magnetic signals from intact Nc molecules. In addition to linear chains, dimers of the Ni-Cp fragments form along the $\langle 1 \bar{1} 0\rangle$ directions, requiring gold adatoms for their creation.  These results demonstrate the feasibility of constructing complex nanostructures based on metallocenes via on-surface synthesis, opening the possibility for realizing low-dimensional magnetic systems by selecting substrates that preserve the magnetic moment of the fragments.
\end{abstract}

\date{\today}

\maketitle

\section{Introduction}

Metallocene molecules are gaining research interest due to their potential applications as catalysts \cite{Kaminsky_2001,Kaminsky_2004, Kaminsky_2007,Shaw_2022,Resconi_2000}, precursors for organo-metallic polymerization \cite{Okuda_2023, Jezequel_2001}, spintronic devices \cite{Ormaza_2017b, Wei_2011, Zhang_2018}, single-spin probes \cite{Ormaza_2017a,czap_probing_2019,Verlhac2019}, and other applications.  A metallocene consists of a transition metal atom with a valence of $+2$, coordinated by two planar cyclopentadienyl (C$_5$H$_5^-$, Cp) ligands, Fig.~\ref{Fig_1} (a). This arrangement provides both structural stability and chemical reactivity to the molecule. In the gas phase, metallocenes are stable and intact at room temperature in an inert atmosphere. In vacuum, they are known to present instabilities on some metallic substrates leading to dissociation and chemical reactions. For instance, ferrocene (FeCp$_2$) deposition on cold Au(111) surface ($<80$ K) results into islands consisting of decomposed fragments~\cite{Braun_2006}, but is instead stable when deposited on Cu(100)~\cite{Ormaza_2015} becoming then amenable for tailoring well-controlled molecular suprastructures~\cite{Ormaza_2016}. Interestingly, metallocene decomposition can be used to create active catalytic sites, fabricate thin films via chemical vapor deposition, and synthesize advanced materials like carbon nanotubes or graphene.

Unlike ferrocene, which is spin-less, nickelocene (NiCp$_2$) has a $S=1$ spin with two low-energy spin excitations arising from an easy-plane magnetic anisotropy of $4$ meV~\cite{Li_1992}. The chemical integrity and magnetic character of nickelocene are usually preserved on cold surfaces. Nickelocene forms ordered arrays on Cu(100) where molecules pair via van der Waals interactions~\cite{Bachellier_2016}. Similar arrays are seen on other metallic substrates such as Pb(111) leading to incommensurability~\cite{Mier_JPCL}. Nickelocene fragmentation is known to occur above 200 K on Ag(100)~\cite{Welipitiya_1998,Pugmire_1999,Pugmire_2001} and Cu(100)~\cite{Bachellier_2016}, although studies at the single-molecule level are still lacking. The magnetic anisotropy is weakly affected by the surface, typically ranging from 3.2 meV on Cu(100) ~\cite{Ormaza_2017a} to 3.9 meV on Pb(111)~\cite{Mier_JPCL}. A recent theoretical study confirmed this ``magnetic" resilience~\cite{Alessio_2023}, also emphasizing the need for further research on small molecular magnetic systems. 

Nickelocene fragments present a promising avenue for further elucidating the origin of this resilience. Their self-assembly on metal surfaces, moreover, offers unique opportunities as versatile building blocks for molecular spintronic devices, spin sensors, and quantum computing elements. Motivated by this potential, we thoroughly investigated the \Nc dissociation by depositing \Nc on a Au(111) substrate held at room temperature, and then imaging the resulting fragments with a low-temperature scanning tunneling microscope (STM). Distinct fragments are identified, some of them forming one-dimensional structures. Through density functional theory (DFT) calculations, we identified the dissociation fragments, characterized the resulting nanostructures, and revealed the forces governing the interaction between \Nc molecules and the Au(111) substrate.

\section{Deposition of N\MakeLowercase{i}C\MakeLowercase{p}$_2$ molecules on A\MakeLowercase{u}(111)}

\begin{figure}
    \centering
    \includegraphics[width = 0.45\textwidth]{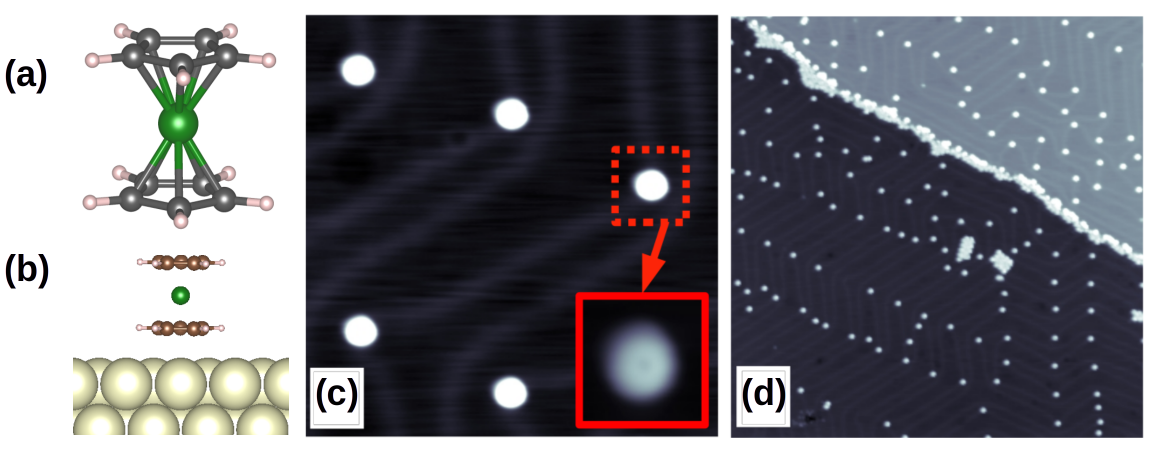}
    \caption{\textbf{Single Nickelocene adsorption on Au(111)}. (a) Ball-and-stick model of a Nickelocene molecule (NiCp$_2$) composed of two cyclopentadienyl rings (Cp) and a Ni atom. (b) Lateral view of an adsorbed Nickelocene molecule on a Au(111) surface. (c) Constant current STM images showing topography of \Nc molecules adsorbed on the surface of Au(111) of an area of $20\times20$ nm$^2$, and imaged for a constant current of  50~pA at a bias of 50~mV, tip grounded.  Preferentially, the  molecules adsorb on herringbone elbow sites. The inset shows a zoomed-in image of an isolated molecule. Both ring-like shape and corrugation  coincide with published images of intact molecules~\cite{Ormaza_2017a}. A larger area is shown in (d), $100\times100$ nm$^2$, taken at 10 pA and 50 mV that shows \Nc decorating crystal step edges, as well as the  formation of small molecular islands and isolated molecules spread all across the surface into herringbone reconstruction sites. }
    \label{Fig_1}
\end{figure}

Constant-current STM images of \Nc molecules deposited on the clean Au (111) surface cooled at 4.2~K shows the corrugation and shape of \Nc molecules with one of the Cp rings facing the surface, Fig.~\ref{Fig_1} (b), as on other surfaces~\cite{Bachellier_2016, Mier_JPCL}. The molecules adsorb intact and individually at low fluxes.  Figure \ref{Fig_1} (c) shows the surface topography scanned at 4.2~K where in the inset shows ring-like shape of isolated NiCp$_{2}$ molecule which agrees well with the above mentioned geometrical description. It can be seen that the molecule preferentially adsorbs at the herringbone elbows, which are well-known active sites for adsorbates. On a larger imaged area, Fig.~\ref{Fig_1} (d), the molecules
decorate Au step edges and furthermore form small ordered arrangements composed of horizontal and vertical molecules as described in Ref.~\cite{Ormaza_2015, Bachellier_2016, Mier_JPCL}.
 Images of isolated NiCp$_{2}$ molecules have an apparent height of 3.5~{\AA}, while the images of molecular monolayers reached 4~{\AA}, similar to \Nc on Cu(100)~\cite{Bachellier_2016}.

\begin{figure}
    \centering
    \includegraphics[width = 0.35\textwidth]{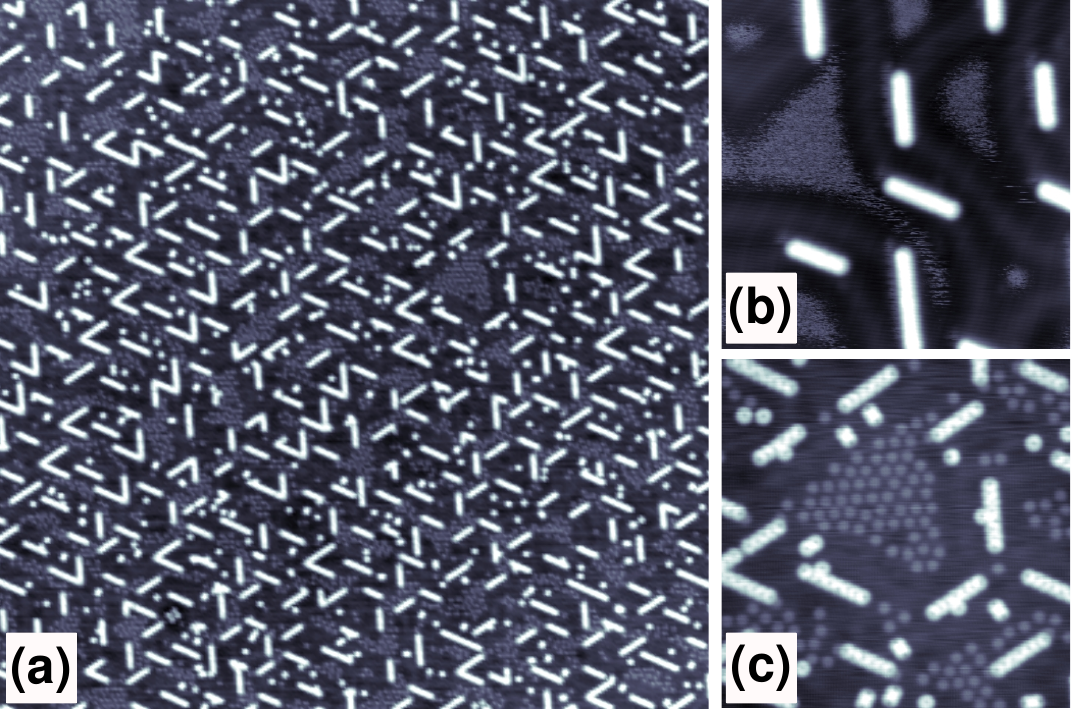}%
     \caption{\textbf{Dissociation of \Nc and new structures on Au(111)}. Constant-current STM images of molecules deposited on a room-temperature Au(111) surface, reveal two distinct types of objects, referred to as type-A and type-B fragments. (a) Image taken at 4.2 K shows type-A fragments forming one-dimensional chains, while type-B fragments arrange in an ordered pattern within the intervening spaces (imaging parameters: $100 \times 100$ nm$^2$, 20 pA, 0.02 V). (b)  Image taken at 77 K where type-A fragments remain well-resolved, whereas type-B fragments exhibit dynamic instability, preventing individual resolution in the topography (imaging parameters: $25 \times 25$ nm$^2$, 20 pA, -1 V). (c) For comparison, an image of an equivalent area at 4.2 K provides a clearer view of the fragment structures (imaging parameters: $20 \times 20$ nm$^2$, 20 pA, 0.02 V). }%
    \label{Fig_2}%
\end{figure}

Molecular deposition on a room-temperature sample results in molecular dissociation. This effect has been observed in previous studies, where sample temperature was shown to significantly influence molecular bond breaking~\cite{Pugmire_2001}. Figure~\ref{Fig_2} presents topographic images of the \Nc fragments  measured at (a, c) 4.2~K and (b) 77~K. At 77~K, two distinct structures are visible: one with low corrugation and a somewhat noisy appearance, and another with higher corrugation, displaying a distinct 1-D structure. The low-corrugation fragments appear featureless and do not seem to aggregate into interacting clusters. When the temperature is lowered to 4.2~K (Fig.~\ref{Fig_2} (c)), both types of structures exhibit more defined features, and the noise is eliminated, suggesting that these structures consist of dissociated components of the original \Nc molecules. The diffusion of the low-corrugation fragments at 77~K likely contributes to the noisy appearance observed in Fig.~\ref{Fig_2} (b).

\begin{figure}[hbt!]
    \centering
    \includegraphics[width = 0.4\textwidth]{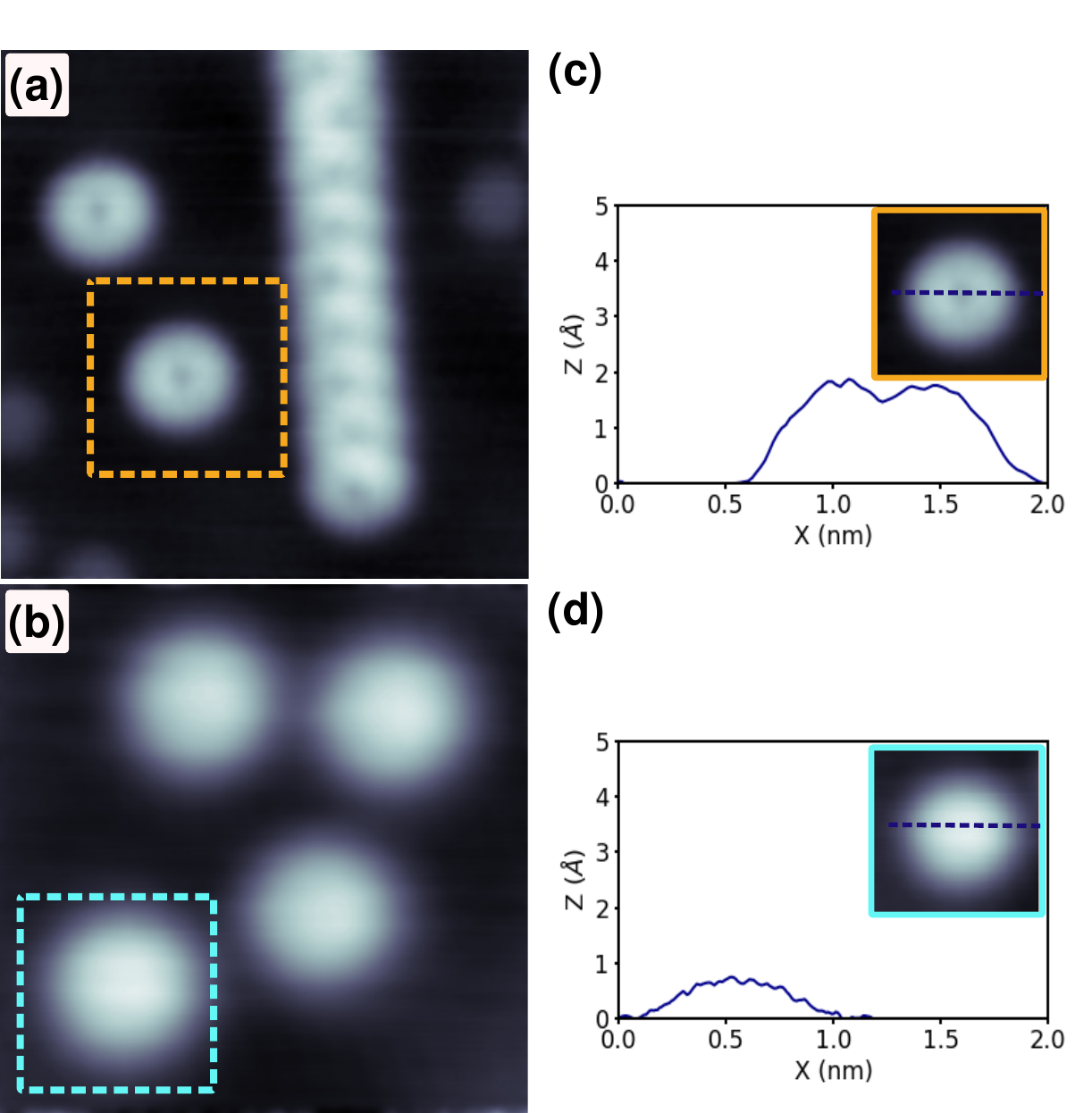}
    \caption{\textbf{Comparison between the two types of fragments}. STM topographic data on the two types of fragments with their corresponding measured height profiles. (a) Type-A fragments found in isolated and chained adsorption state (imaging parameters: $5{\times}5$ nm$^2$, 20~pA, 20~mV). (b) Four type-B fragments with the average nearest-nerighbor distance of $8.7$~\AA~ (imaging parameters: $2.33\times2.33$~nm$^2$, $20$~pA, $20$~mV). (c) and (d) show the profile along the line plotted in the insets of the respective plots.}
    \label{Fig_3}
\end{figure}

 We designate the ring-shaped fragments that form one-dimensional chain structures as type-A, while the fragments exhibiting smaller corrugation and an ordered arrangement are labeled as type-B. Figure~\ref{Fig_3} (a) displays type-A fragments both in isolated form and as part of a one-dimensional chain. Panel (c) provides the lateral profile of an individual type-A fragment, characterized by a symmetric ring shape with a maximum corrugation of $1.70{\pm}0.05$~{\AA}, ruling out the possibility that these are intact molecules (expected corrugation 4~{\AA}; see discussion above). 
 The one-dimensional structures are formed by ring-like structures with a small depression in each ring, but this time the depression is not centered and the rings are not symmetrical, reminiscent of constant-current STM images of tilted \Nc molecules~\cite{Verlhac2019}.
 The spacing between rings is approximately $5.1\pm0.1$~{\AA}, aligning with the Au atomic distance along the [$11\bar{2}]$ direction, consistent with chain growth along the FCC domains of the herringbone reconstruction, as visible in Fig.~\ref{Fig_2} \lolo{(b)} and corroborated by recent studies on molecular growth and diffusion~\cite{Edmondson_2022}. The corrugation of the chain is $1.9{\pm}0.1$~{\AA}. 

Type-B fragments exhibit round shapes with lower corrugation, as shown in Fig.~\ref{Fig_3} (b). Each type-B fragment has a corrugation of $0.6{\pm}0.1$~{\AA}, as illustrated in Fig.~\ref{Fig_3} (d). These fragments assemble into superlattices as shown in Fig.~\ref{Fig_2} \lolo{(c)}. In these ordered structures, the average distance between fragments is $8.7{\pm}0.1$~{\AA}, aligning closely with three lattice parameters {of Au(111)} (8.65~\AA) along the $\langle 1\bar{1}0\rangle$ directions and matching the triangular symmetry of the  surface. Originating at the herringbone elbows, these aggregates modify the reconstruction, creating a large triangular region reminiscent of Ni atom adsorption on Au(111) reported in Ref.~\cite{Meyer_1996}. Unlike the substitutional Ni adatoms in Ref.~\cite{Meyer_1996}, the observed structures here display the above positive corrugation.

The superlattice of type-B fragments are intriguingly regular.
Regular patterns of adsorbates have been observed on noble metal surfaces, often attributed to  a surface state mediated interaction~\cite{Repp_2000}. The adsorbate spacing is approximately half the Fermi wavelength of the surface, \( \lambda_{F}/2 \), which for Au(111) is \( \lambda_{F}/2 = 18 \)~\AA, see for example Ref.~\cite{Sotthewes}. In contrast, we find that the type-B fragments align along the \( \langle 1\bar{1}0 \rangle \) directions and are spaced at intervals of 8.7~\AA, or three surface lattice parameters.



The interaction of the fragmented adsorbates with the Au(111) surface is sufficiently strong to induce a partial reconstruction of the herringbone structure. In Fig.~\ref{Fig_2} \lolo{(b)}, type-A fragments are observed to align along the FCC terraces, contributing to alterations in the herringbone arrangement. In contrast, type-B fragments occupy the larger areas around the elbows and regions preserved in the remaining herringbone structure. At higher molecular coverages the herringbone reconstruction disappears and triangular patterns start forming because of the ubiquitous growth of chains of the type-A fragments.

\subsection{Identification of fragments}

To identify the fragments, we conducted DFT calculations on various possible fragment configurations and structures, comparing the computed STM images with experimental observations. These calculations further enable us to explore the geometry, electronic structure, and overall properties of each structure, providing a comprehensive view of the products resulting from \Nc dissociation on the Au(111) surface.
\\

\begin{figure}[t!]
    \centering
    \includegraphics[width = \columnwidth]{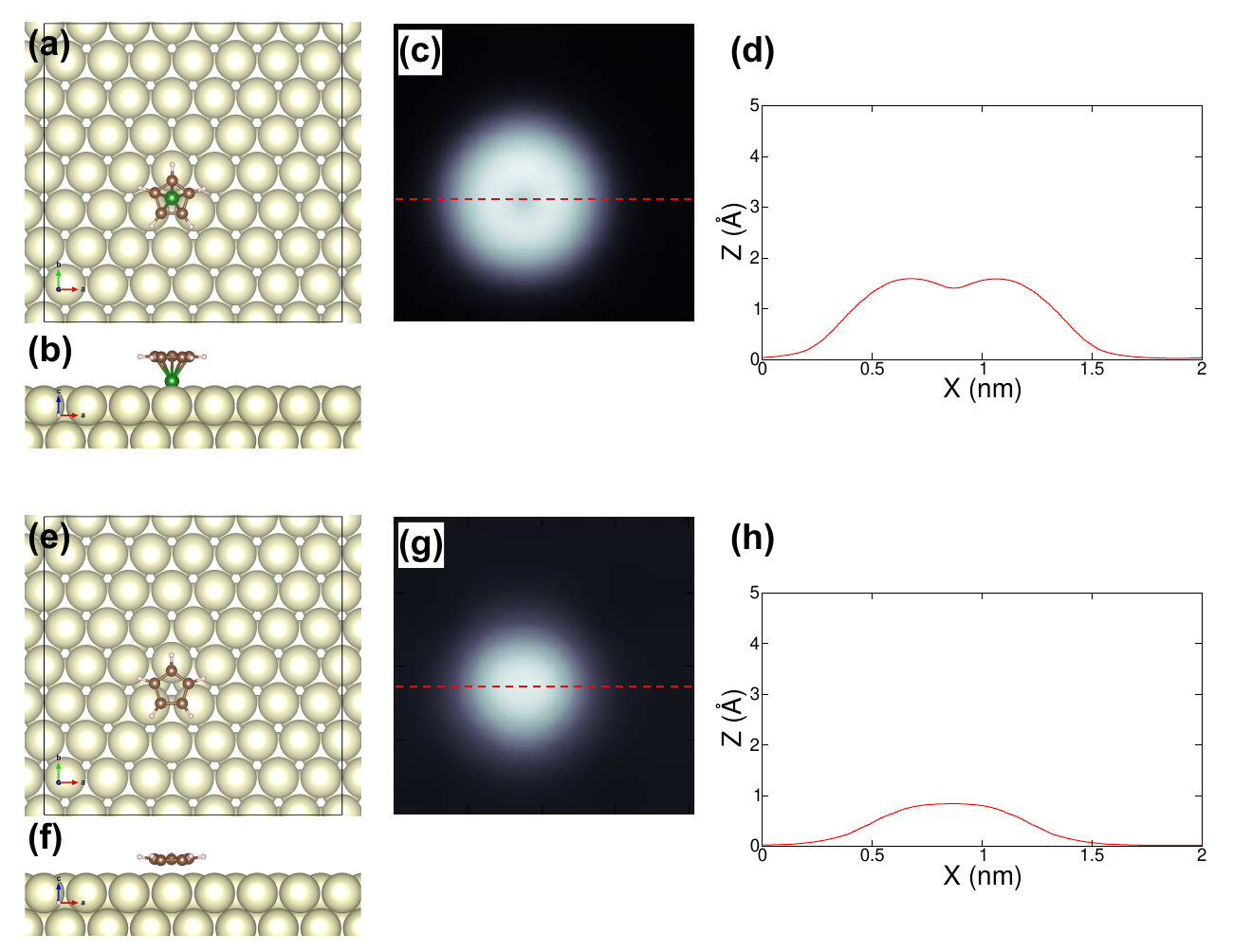}
    \caption{{\textbf{Computed results of the two fragment types}. (a) Structure of the energetically most stable fragment, consisting of a Ni atom on an FCC hollow site of an FCC Au(111) terrace, capped by a Cp ring. The scheme corresponds to the unit cell of the fully-relaxed DFT calculation. (b) Lateral view of the same configuration, with the slab and vacuum depth truncated for clarity. (c) Computed STM image of the fragment, accompanied by its height profile in (d). The computed STM image aligns well with the experimental STM image of the type-A fragment. (e) and (f) show the structure of a Cp ring on the surface. (g) and (h) display the corresponding STM image and profile data. These computed results are in good agreement with the data for type-B fragments, although similar results are obtained for single non-substitutional Ni adatoms. }}
    \label{Fig_4}%
\end{figure}

\begin{figure}[hbt!]
    \includegraphics[width = 0.35\textwidth]{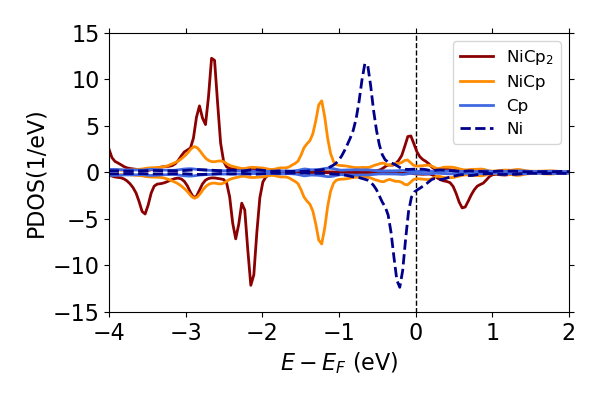}
    \caption{\textbf{Electronic structure of the adsorbates}. Calculated projected density of states (PDOS) for the \Nc molecule and fragments NiCp, Cp and Ni on Au(111) surface. Positive values refer to the PDOS on spin-up states and negative values on spin-down states. The PDOS on \Nc molecular states shows the stronger spin polarization with substantial emptying of the $\pi_d$ LUMO orbitals due to the interaction with the substrate. The PDOS on Ni adsorbates show some spin polarization, and the PDOS on the fragments NiCp and Cp are not spin polarized. The width of the PDOS peaks also give information on the strength of the interaction with the surface, showing that NiCp present very broad PDOS and the Cp PDOS is basically flat due to the very strong interaction of these radical species with the substrate.}
    \label{Fig_5}
\end{figure}

Figure~\ref{Fig_4} presents two possible fragment configurations that align with the experimental observations. The adsorbed structure depicted in Fig.~\ref{Fig_4} (a) and (b) yields the computed STM image in (c) with a profile in (d), showing quantitative agreement with the experimental data for a single type-A fragment (refer to Fig.~\ref{Fig_3} (a) and (c)). This fragment adsorbs with a chemisorption energy of \( E_B = 4.100 \) eV, defined as the energy required to desorb the NiCp fragment from the surface:
\[
E_B = E_{\text{NiCp}} + E_{\text{Au(111)}} - E_{\text{NiCp/Au(111)}}.
\]
In contrast, the full \Nc molecule exhibits a lower binding energy, \( E_B = 1.370 \) eV, primarily driven by van der Waals interactions with the surface. Figure~\ref{Fig_5} displays the density of states (DOS) for various systems, projected onto the molecular electronic structure of each system (PDOS). The PDOS for \Nd, reveals two distinct peaks around the Fermi energy (set as the zero energy reference), corresponding to the two \( d_\pi \) orbitals of the molecule with significant contributions from Ni \( d \)-orbitals. In agreement with the spin \( S = 1 \) of \Nd, these two orbitals are singly occupied, resulting in peaks positioned symmetrically above and below the Fermi level~\cite{Ormaza_2017a, Mier_JPCL}. However, upon adsorption on Au(111), the occupied peak shifts towards the Fermi energy, indicating substantial charge transfer from the molecule to the substrate and a consequent loss of magnetic moment.

Figure~\ref{Fig_5} also displays the PDOS of the NiCp fragment, which we associate with the type-A fragment. Here, we observe identical contributions from both spin channels, indicating that the system is unpolarized. Furthermore, the PDOS is broad around the Fermi energy, suggesting significant charge redistribution and strong coupling with the surface electronic states. Consequently, this fragment exhibits enhanced reactivity and strong adsorption on the surface. 

Bader charge analysis reveals that the fragment donates approximately 0.4 electrons to the surface, consistent with a Ni(II) oxidation state. The preferred adsorption site positions the Ni atom in an FCC hollow site on the Au(111) FCC terrace. Despite the Ni(II) state, the strong electronic hybridization with the surface results in a complete loss of spin polarization for the fragment, with the calculations indicating zero magnetic moment. This finding aligns with the absence of magnetic excitations in the experimental \( dI/dV \) spectra recorded for individual type-A fragments.

When the Cp ring, rather than the Ni atom, faces the surface, the binding energy decreases to \( E_B = 1.278 \) eV, closely resembling that of the full \Nc molecule, indicating that the interaction is primarily van der Waals in nature. In this configuration, the magnetic moment remains intact at \( 1.14~\mu_B \) due to the minimal hybridization with the surface. According to our DFT calculations, the simulated STM image of this configuration exhibits a corrugation of 4.1~{\AA}, similar to that of the full \Nc molecule, despite the shorter length of the fragment. Consequently, this configuration cannot correspond to either the type-A or type-B fragments.

Figure~\ref{Fig_4} (e), (f), (g), and (h) show a single Cp ring adsorbed on the surface. The binding energy is \( E_B = 1.772 \) eV, which is higher than that of the full \Nc molecule, reflecting the radical nature of the Cp ring. Of this binding energy, 0.796 eV stems from van der Waals interactions, while the remaining energy arises from charge redistribution between the ring and the substrate. This redistribution is confirmed by the computed induced electronic charge, indicating a substantial rearrangement of charge between the surface and the Cp ring. Additionally, the PDOS in Fig.~\ref{Fig_5} is broad and low, lacking clear resonant features, further demonstrating that the electronic structure of the Cp ring is ill-defined due to extensive mixing with the substrate’s electronic states.

The computed corrugation of the STM image,  Fig.~\ref{Fig_4} (h), is 0.8~{\AA}, closer to the experimental corrugation of type-B fragments. Additionally, the constant current image is very much featureless like the experimental one, Fig.~\ref{Fig_3} (b). Similar to the Cp fragment, a Ni adatom yields an STM-image corrugation of 0.9~{\AA}. The only difference between the two STM images is the FWHM of the image, 7~{\AA} of the Cp fragment, while it is 6~{\AA} for the Ni adatom. This difference is difficult to verify in experimental STM images.

Ni adatoms on Au(111) retain a sizable magnetic moment of $0.96~\mu_B$, which agrees with the tail of $d$-electrons spreading above the Fermi energy in Fig.~\ref{Fig_5}. However, no magnetic excitations due to the adatom's spin flip is expected because a $S=1/2$ system has no anisotropy. Thus the absence of magnetic excitations in the $dI/dV$ spectra does not allow us to conclude on the nature of the type-B fragment.

The aggregates of type-B fragments observed near the herringbone elbows in Fig.~\ref{Fig_2} suggest that Ni adatoms are unlikely to be the type-B fragments. When Ni adatoms adsorb on Au(111), they typically replace surface gold atoms, forming compact clusters of Ni atoms~\cite{Meyer_1996}, which contrasts with the structures we observe. In our findings, the fragments do not cluster closely but instead maintain a certain distance from each other, likely indicating electrostatic repulsion~\cite{Fernandez-Torrente_2007}. Considering the radical character of Cp rings, we interpret these type-B aggregates as likely composed of Cp rings, which are expected to be in excess due to the association of Ni atoms with type-A fragments.

 In conclusion, type-A fragments are likely NiCp molecules, while type-B fragments correspond to the Cp rings. Their distinct compositions result in markedly different adsorption patterns: type-A fragments form 1-D arrays aligned with the herringbone structure of the substrate, while the Cp rings (type-B fragments) aggregate along the high-symmetry \( \langle 1\bar{1}0 \rangle \) directions of the surface, maintaining an equidistant network. This arrangement is likely due to a balance between attractive van der Waals forces and repulsive electrostatic interactions.
 
\section{Dimers of fragments}

\begin{figure}[t!]
    \centering
    \includegraphics[width = 0.45 \textwidth]{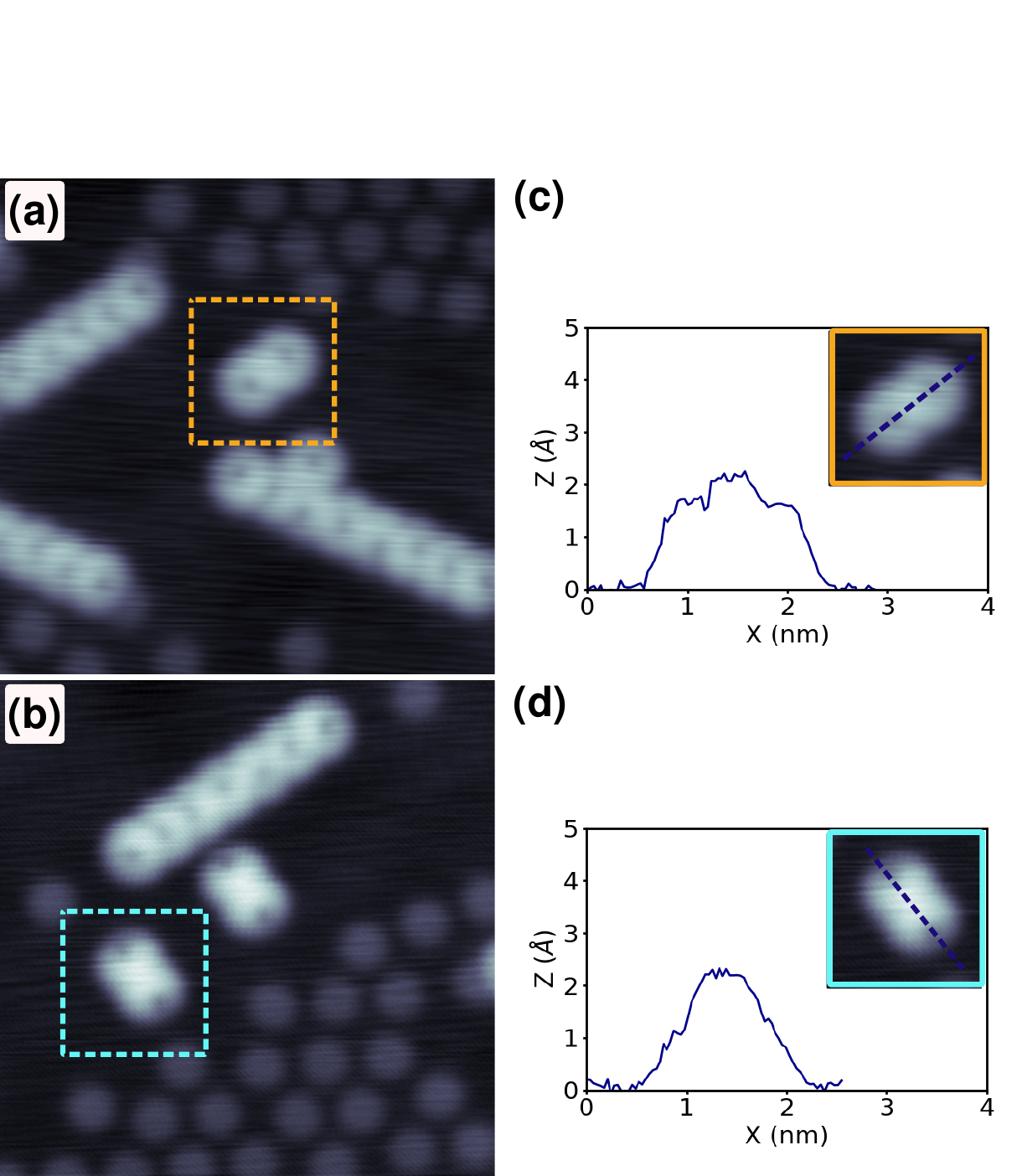}
    \caption{\textbf{Types of fragment dimer.} Constant current STM images showing the formation of different dimer configurations composed of type-A fragments on the Au(111) surface. (a) Dimer aligned along the $[11\bar{2}]$ direction (image parameters: $7\times7$ nm$^2$, 50~pA, -20~mV). (b) Dimer aligned along the $[1\bar{1}0]$ direction (image parameters: $7\times7$ nm$^2$, 20~pA, 20~mV). Panels (c) and (d) display the height profiles of the STM images along the dashed lines shown in the insets.}
    \label{Fig_6}
\end{figure}

\begin{figure}[h!]
    \centering
    \includegraphics[width = 0.45\textwidth]{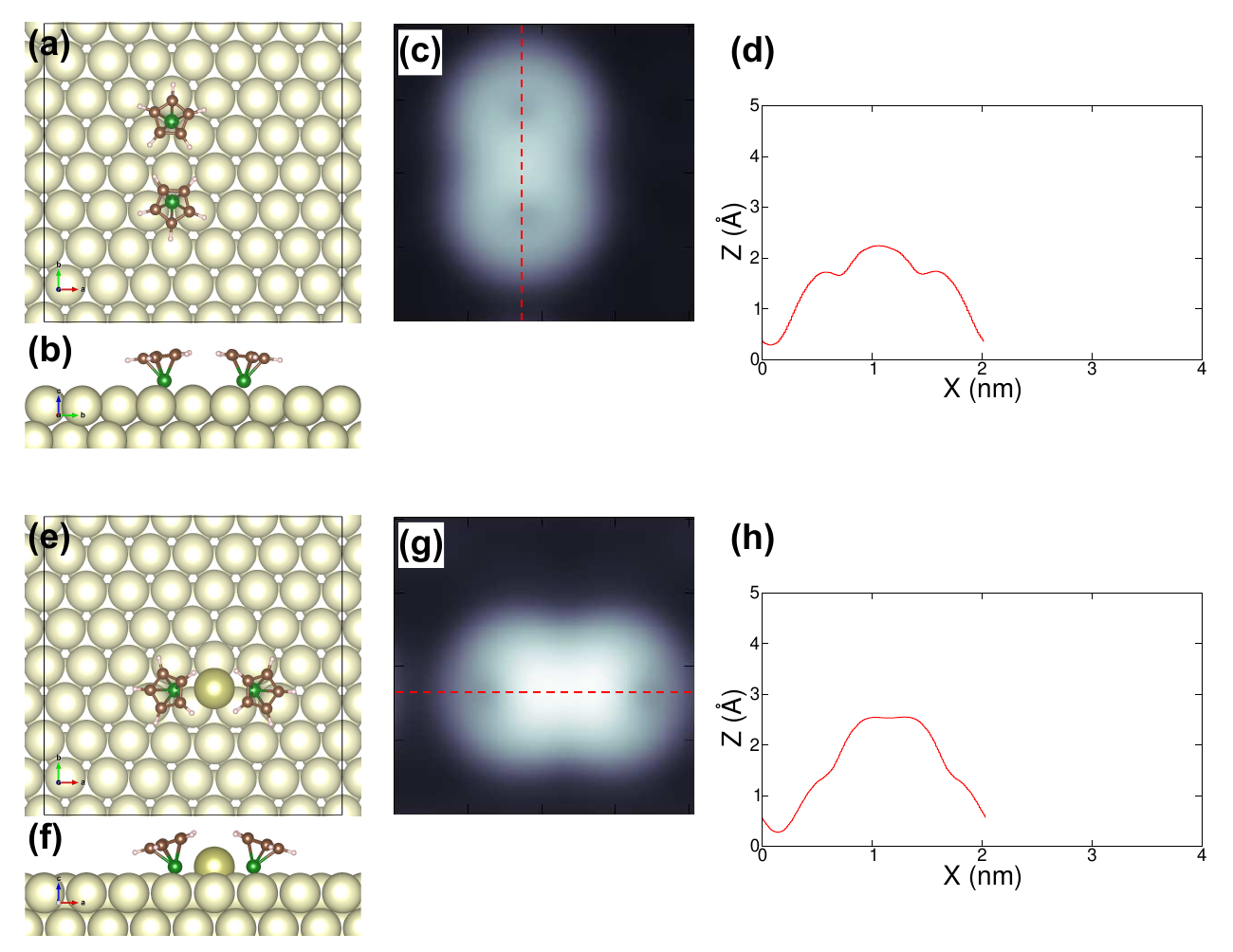}
    \caption{\textbf{DFT Calculations of Dimers}. DFT results for a dimer aligned along the  $[11\bar{2}]$ direction, showing the adsorption geometry in (a) and (b), and the corresponding simulated STM image in (c), with the profile along the dimer axis plotted in (d). For a dimer aligned along the $[1\bar{1}0]$ direction, the presence of a gold adatom stabilizes the dimer structure, yielding excellent agreement with the experimentally measured STM images, as shown in (e)–(h). }%
    \label{Fig_7}%
\end{figure}

Interestingly, STM images reveal two types of dimers formed by type-A fragments, as shown in Fig.~\ref{Fig_6}. In contrast, type-B fragments do not nucleate or form dimers consistent with our previous analysis. Type-A fragments, however, form dimers along the $\langle 11\bar{2}\rangle$ directions [Fig.~\ref{Fig_6} (a)] and the $\langle 1\bar{1}0 \rangle$ directions [Fig.~\ref{Fig_6} (b)]. This observation is particularly significant given that longer chains of type-A fragments are observed to grow exclusively along the $\langle 11\bar{2}\rangle$ directions.

DFT calculations indicate that type-A fragments readily form dimers when the Ni adatom occupies FCC hollow sites along the $\langle 11\bar{2}\rangle$ directions. Figures~\ref{Fig_7} (a) and (b) show DFT results for a dimer configuration that matches the inter-fragment distance observed experimentally in Fig.~\ref{Fig_6} (b). The steric repulsion between hydrogen atoms induces a molecular tilt, which is essential for reproducing the features seen in the experimental STM images, as illustrated in Fig.~\ref{Fig_7} (c) and (d).

Reproducing the dimers along the $\langle 1\bar{1}0 \rangle$ directions proved considerably more challenging, as these dimers represent the shortest possible distance between fragments. Along the $\langle 1\bar{1}0 \rangle$ directions, this configuration requires an FCC hollow site positioned between the Ni atoms to mitigate steric hindrance. However, under these conditions, the computed dimer image does not match the experimental data. For Au(111) surfaces, previous studies (e.g., \cite{Mielke_2014, Robles_2020, Berger_2022} and references therein) have shown that certain adsorbates can trap gold adatoms from the substrate. By introducing a gold adatom into the hollow site between the two type-A fragments, we achieve excellent agreement with the experimental images, as shown in Fig.~\ref{Fig_7} (e–h). The presence of the gold adatom provides the necessary steric bulk to stabilize the fragments in a symmetric configuration, which is reflected in the computed STM image.

The binding energies per fragment reveal how the dimers are created. For the $\langle 11\bar{2}\rangle$ directions, the dimers show a binding energy $E_B=4.078$ eV, very close to the single fragment ($E_B=4.100$ eV). The dimer is actually slightly less binding than the single fragment because of the fragment-fragment steric repulsion revealed above. For the $\langle1\bar{1}0\rangle$ directions, the gold adatom actually contributes positively to the fragment dimer and the binding energy is more stable, $E_B=4.322$ eV per fragment. The van der Waals contribution to the interaction is the main binding force between fragments.

\section{Chains of fragments}

The chains are formed by type-A fragments that grow along the $\langle11\bar{2}\rangle$ family of directions on the surface. This is clearly seen in Fig.~\ref{Fig_2} (b) where different chains are shown following the $\langle11\bar{2}\rangle$ directions and thus forming angles that are multiples of $60^\circ$. The chains do not grow along the $\langle\bar{1}10\rangle$ directions despite the fact that we find dimers that are aligned along the $\langle\bar{1}10\rangle$ directions. Moreover, even if the chains grow along the $\langle11\bar{2}\rangle$ directions, they rarely reach beyond 10 fragments.

\begin{figure}[t!]
    \centering
    \includegraphics[width = \columnwidth]{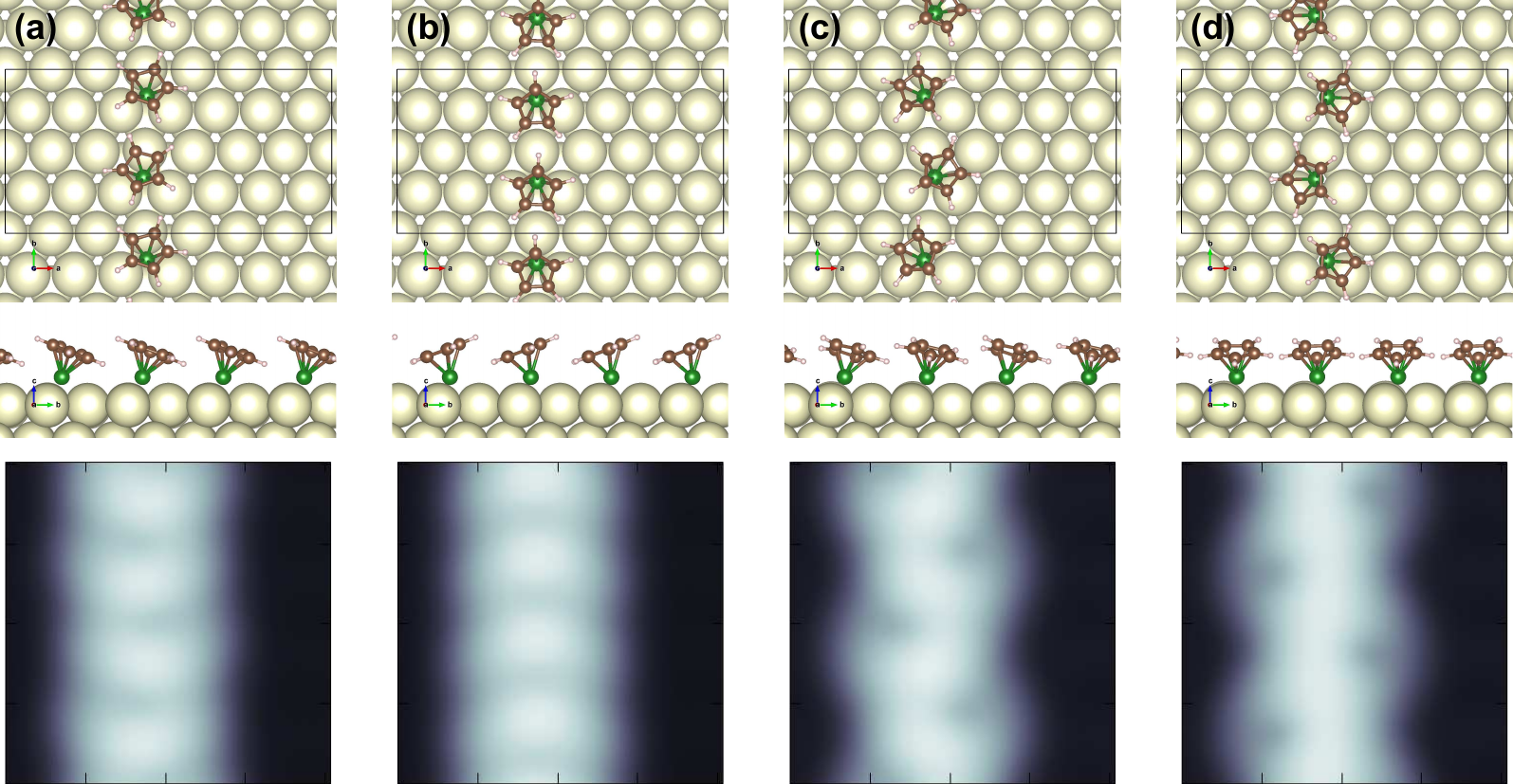} 
    \caption{\textbf{Fragment chain configurations}. DFT results for a chain formed by periodically repeated dimers of type-A fragments along the $\langle 11\bar{2} \rangle$ direction. The top panels show the geometry of the fragments, which adjust to minimize steric repulsion between molecular protons, resulting in distinctive fragment patterns within the chain. The effect of these patterns is evident in the lower panel, where the computed STM images are displayed. These images provide insight into the experimentally observed chiral patterns of the chain.}%
    \label{Fig_8}%
\end{figure}

To explain these observations, we note that the primary interaction driving chain formation is the Ni-substrate interaction. Specifically, the Ni atom coordinates with an FCC hollow site on the surface, with the molecule aligning itself to minimize steric hindrance. This naturally results in the formation of 1-D structures. Along the $\langle 1\bar{1}0 \rangle$ directions, however, an additional adatom would be required, implying that an ample supply of adatoms is necessary for fragments to capture them. While certain thermodynamic conditions may enable this, it would likely result in clusters of fragments rather than 1-D chains. Under the conditions of our experiment, the scarcity of adatoms is probably the limiting factor, preventing the formation of more complex structures.

We performed DFT calculations using a dimer unit cell to model an infinite chain, produced by periodically repeating the dimer. Steric hindrance plays a significant role in lowering the energy of the chain and generating a distinct pattern. Figure~\ref{Fig_8} displays both the atomic structure of these chains and the corresponding computed STM images, where we successfully reproduce the chiral patterns observed in Figs.~\ref{Fig_2} and \ref{Fig_3}. Our calculations indicate that the lowest-energy chain configuration is the one shown in Fig.~\ref{Fig_8} (a), with an energy reduction of 81 meV compared to other chain configurations. This result is noteworthy, as it suggests that neither the configuration with the furthest-separated hydrogen atoms nor the non-chiral arrangement represents the minimum energy state.

\begin{figure}
    \centering
    \includegraphics[width = 0.35\textwidth]{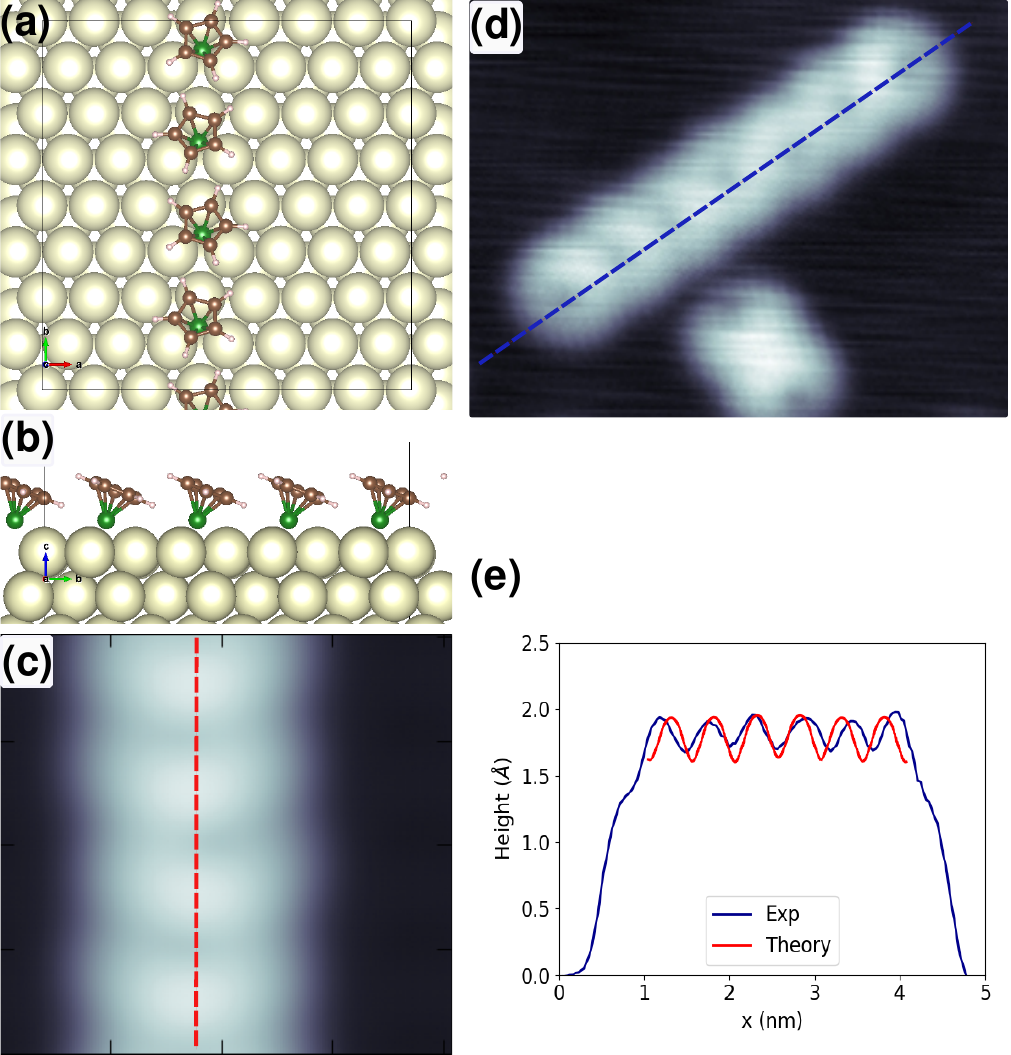}
    \caption{\textbf{Structure and corrugation of the fragment chain}. (a, b) Top and lateral views of a four-fragment unit cell scheme, periodically repeated to simulate a type-A fragment chain. (c) Computed STM image, with a dashed line indicating the profile line. (d) Experimental STM image of a type-A fragment chain, with a line marking the corrugation profile ($3.3 \times 4.4$ nm$^2$, 20~pA, 20~mV). (e) Comparison of experimental and theoretical profiles, showing good agreement. The corrugation is nearly identical, with the only difference being the origin of heights between theory and experiment; here, the theoretical curve has been shifted vertically for comparison since it corresponds to an infinite case, while the experimental one is finite. The finite nature of the experimental chain results in increased inter-fragment distances near the chain edges due to intrinsic chain strain. }
    \label{Fig_9}
\end{figure}

However, these results may be influenced by the constraints of the small unit cell. To address this, we performed additional calculations using a doubled unit cell. Figure~\ref{Fig_9} presents the results for a unit cell containing two dimers, allowing for direct comparison with experimental data. We observe that the smooth chiral pattern remains the lowest-energy configuration, displaying excellent agreement with the experimental STM image. Figure~\ref{Fig_9} also compares the STM image profiles from experiment and theory, showing that the calculated corrugation and periodicity align well with the experimental observations. Nonetheless, the experiment involves a finite chain, where strain causes an increasing separation between fragments as we move outward from the chain’s center. Additionally, we find that individual fragments can be manipulated out of the chain without disrupting the overall structure, consistent with our calculations indicating weak inter-fragment interactions.

\section{Conclusions}
We have investigated the deposition of \Nc molecules on Au(111) at different substrate temperatures. At low temperature (4.2~K), \Nc molecules adsorb individually at herringbone elbows and step edges, progressively forming ordered patterns as molecular flux increases. At room temperature, the {molecular flux produces} a markedly different adsorption patterns, indicating that \Nc molecules dissociate at this temperature. By combining STM imaging with DFT calculations, we identified two primary types of fragments, although the presence of some Ni adatoms cannot be entirely excluded. 

The first fragment type consists of a Ni atom capped by a Cp ring, with the Ni atom adsorbed on an FCC hollow site of the Au(111) FCC terraces. These fragments appear more protruding in STM images and form distinct 1-D structures along the $\langle 11\bar{2}\rangle$ directions of the surface. The second fragment type aggregates while maintaining a certain distance, suggesting short-range repulsive interactions. We attribute these to Cp rings, which, being radical molecules, interact strongly with the substrate yet have low diffusion barriers. Both experimental data and theoretical calculations indicate that neither fragment type exhibits magnetism, in contrast to the magnetic signals observed in intact adsorbed \Nc molecules.

Fragments are also found in dimer configurations, and, unlike the chains, these dimers can form along both $\langle 1 \bar{1}0\rangle$ and $\langle 11\bar{2}\rangle$ directions. We demonstrate that dimers along $\langle 1 \bar{1}0\rangle$ require the presence of a gold adatom for stabilization, possibly explaining the absence of chains in this direction due to the scarcity of adatoms.

Both dimers and chains exhibit strain, and the steric hindrance of Cp protons introduces chiral patterns in the chains. Our DFT calculations successfully capture these effects, showing good agreement with experimental observations.

These findings suggest the potential for creating complex nanostructures based on metallocenes through on-surface synthesis, leveraging STM manipulation capabilities and the thermodynamic properties of metallocene fragments. An exciting avenue for future research would be to realize these 1-D chains on substrates that do not quench the magnetic moment of the fragments, potentially enabling the study of low-dimensional magnetic systems.

\section*{Methods}

A single-crystal Au(111) surface was cleaned in situ through repeated cycles of Ar$^{+}$ sputtering and annealing up to 800~K, at a base pressure of $2 \times 10^{-10}$ mbar, until a clean surface with herringbone reconstruction was obtained. Two types of samples were prepared, varying the deposition process of \Nc molecules on the surface.

For the first preparation, the Au crystal was thermally stabilized at liquid helium temperature. Once the sample temperature reached 4.2~K, it was removed from the cooling stage and held on a wobble stick for deposition of NiCp$_2$ molecules nearby. Using the pressure differential between chambers within the high-vacuum system, NiCp$_2$ was deposited onto the Au surface for approximately 2–3 seconds. {The sample was then re-inserted in the cooling stage. Overall, the deposition procedure takes less than a minute, thus the sample remains below $100$~K.} 

For the second sample, a clean Au crystal was thermally stabilized at room temperature, and NiCp$_2$ molecules were deposited on the surface for about 40 seconds. The sample was then transferred to the STM chamber, where measurements were conducted at 4.2~K.

Density functional theory (DFT) calculations were performed using the VASP code \cite{kresse_efficiency_1996}. The projector augmented-wave (PAW) method was applied to treat the core electrons \cite{kresse_ultrasoft_1999}, with wave functions expanded in a plane-wave basis set using an energy cutoff of 500~eV. For exchange and correlation, we employed the PBE \cite{perdew_generalized_1996} flavor of the generalized gradient approximation (GGA). The missing van der Waals interactions in this functional were accounted for using the Tkatchenko-Scheffler model \cite{tkatchenko_accurate_2009}.

The Au(111) surface was modeled via the slab method, consisting of four Au layers with a vacuum layer of at least 20~\AA. The geometry was relaxed for all atoms, except for the two bottom Au layers, until the forces were less than 0.01~eV/\AA. STM images were simulated using the Tersoff-Hamann approximation \cite{tersoff_theory_1985}, following the approach of Bocquet \textit{et al.} \cite{bocquet_theory_2009} as implemented in STMpw \cite{lorente_stmpw_2019}. All STM images were evaluated at a finite bias corresponding to unoccupied states ($+1$~V). Atomic schemes were generated using the VESTA program \cite{momma_vesta_2011}.

\section*{Acknowledgement}

We thank projects  PID2021-127917NB-I00 by MCIN/AEI/10.13039/501100011033, QUAN-000021-01 by Gipuzkoa Provincial Council, IT-1527-22 by Basque Government, 202260I187 by CSIC, ESiM project 101046364 by EU. Views and opinions expressed are however those of the author(s) only and do not necessarily reflect those of the EU. Neither the EU nor the granting authority can be held responsible for them. The Strasbourg authors acknowledge support from the EU’s Horizon 2020 research and innovation programme under the Marie Skłodowska-Curie grant 847471, from the International Center for Frontier Research in Chemistry (Strasbourg), and from project ANR-23-CE09-0036 funded by the ANR.

\bibliography{references}

\begin{thebibliography}{39}%
\makeatletter
\providecommand \@ifxundefined [1]{%
 \@ifx{#1\undefined}
}%
\providecommand \@ifnum [1]{%
 \ifnum #1\expandafter \@firstoftwo
 \else \expandafter \@secondoftwo
 \fi
}%
\providecommand \@ifx [1]{%
 \ifx #1\expandafter \@firstoftwo
 \else \expandafter \@secondoftwo
 \fi
}%
\providecommand \natexlab [1]{#1}%
\providecommand \enquote  [1]{``#1''}%
\providecommand \bibnamefont  [1]{#1}%
\providecommand \bibfnamefont [1]{#1}%
\providecommand \citenamefont [1]{#1}%
\providecommand \href@noop [0]{\@secondoftwo}%
\providecommand \href [0]{\begingroup \@sanitize@url \@href}%
\providecommand \@href[1]{\@@startlink{#1}\@@href}%
\providecommand \@@href[1]{\endgroup#1\@@endlink}%
\providecommand \@sanitize@url [0]{\catcode `\\12\catcode `\$12\catcode
  `\&12\catcode `\#12\catcode `\^12\catcode `\_12\catcode `\%12\relax}%
\providecommand \@@startlink[1]{}%
\providecommand \@@endlink[0]{}%
\providecommand \url  [0]{\begingroup\@sanitize@url \@url }%
\providecommand \@url [1]{\endgroup\@href {#1}{\urlprefix }}%
\providecommand \urlprefix  [0]{URL }%
\providecommand \Eprint [0]{\href }%
\providecommand \doibase [0]{http://dx.doi.org/}%
\providecommand \selectlanguage [0]{\@gobble}%
\providecommand \bibinfo  [0]{\@secondoftwo}%
\providecommand \bibfield  [0]{\@secondoftwo}%
\providecommand \translation [1]{[#1]}%
\providecommand \BibitemOpen [0]{}%
\providecommand \bibitemStop [0]{}%
\providecommand \bibitemNoStop [0]{.\EOS\space}%
\providecommand \EOS [0]{\spacefactor3000\relax}%
\providecommand \BibitemShut  [1]{\csname bibitem#1\endcsname}%
\let\auto@bib@innerbib\@empty
\bibitem [{\citenamefont {Kaminsky}\ and\ \citenamefont
  {Laban}(2001)}]{Kaminsky_2001}%
  \BibitemOpen
  \bibfield  {author} {\bibinfo {author} {\bibfnamefont {W.}~\bibnamefont
  {Kaminsky}}\ and\ \bibinfo {author} {\bibfnamefont {A.}~\bibnamefont
  {Laban}},\ }\href {\doibase https://doi.org/10.1016/S0926-860X(01)00829-8}
  {\bibfield  {journal} {\bibinfo  {journal} {Applied Catalysis A: General}\
  }\textbf {\bibinfo {volume} {222}},\ \bibinfo {pages} {47} (\bibinfo {year}
  {2001})},\ \bibinfo {note} {celebration Issue}\BibitemShut {NoStop}%
\bibitem [{\citenamefont {Kaminsky}(2004)}]{Kaminsky_2004}%
  \BibitemOpen
  \bibfield  {author} {\bibinfo {author} {\bibfnamefont {W.}~\bibnamefont
  {Kaminsky}},\ }\href {\doibase https://doi.org/10.1002/pola.20292} {\bibfield
   {journal} {\bibinfo  {journal} {Journal of Polymer Science Part A: Polymer
  Chemistry}\ }\textbf {\bibinfo {volume} {42}},\ \bibinfo {pages} {3911}
  (\bibinfo {year} {2004})},\ \Eprint
  {http://arxiv.org/abs/https://onlinelibrary.wiley.com/doi/pdf/10.1002/pola.20292}
  {https://onlinelibrary.wiley.com/doi/pdf/10.1002/pola.20292} \BibitemShut
  {NoStop}%
\bibitem [{\citenamefont {Kaminsky}\ and\ \citenamefont
  {Arndt}(2007)}]{Kaminsky_2007}%
  \BibitemOpen
  \bibfield  {author} {\bibinfo {author} {\bibfnamefont {W.}~\bibnamefont
  {Kaminsky}}\ and\ \bibinfo {author} {\bibfnamefont {M.}~\bibnamefont
  {Arndt}},\ }\href {\doibase 10.1007/BFb0103631} {\bibfield  {journal}
  {\bibinfo  {journal} {Adv. Polym. Sci.}\ }\textbf {\bibinfo {volume} {127}},\
  \bibinfo {pages} {143} (\bibinfo {year} {2007})}\BibitemShut {NoStop}%
\bibitem [{\citenamefont {Shaw}\ \emph {et~al.}(2022)\citenamefont {Shaw},
  \citenamefont {Bates}, \citenamefont {Jones},\ and\ \citenamefont
  {Ward}}]{Shaw_2022}%
  \BibitemOpen
  \bibfield  {author} {\bibinfo {author} {\bibfnamefont {M.~S.}\ \bibnamefont
  {Shaw}}, \bibinfo {author} {\bibfnamefont {M.~R.}\ \bibnamefont {Bates}},
  \bibinfo {author} {\bibfnamefont {M.~D.}\ \bibnamefont {Jones}}, \ and\
  \bibinfo {author} {\bibfnamefont {B.~D.}\ \bibnamefont {Ward}},\ }\href
  {\doibase 10.1039/D2PY00335J} {\bibfield  {journal} {\bibinfo  {journal}
  {Polym. Chem.}\ }\textbf {\bibinfo {volume} {13}},\ \bibinfo {pages} {3315}
  (\bibinfo {year} {2022})}\BibitemShut {NoStop}%
\bibitem [{\citenamefont {Resconi}\ \emph {et~al.}(2000)\citenamefont
  {Resconi}, \citenamefont {Cavallo}, \citenamefont {Fait},\ and\ \citenamefont
  {Piemontesi}}]{Resconi_2000}%
  \BibitemOpen
  \bibfield  {author} {\bibinfo {author} {\bibfnamefont {L.}~\bibnamefont
  {Resconi}}, \bibinfo {author} {\bibfnamefont {L.}~\bibnamefont {Cavallo}},
  \bibinfo {author} {\bibfnamefont {A.}~\bibnamefont {Fait}}, \ and\ \bibinfo
  {author} {\bibfnamefont {F.}~\bibnamefont {Piemontesi}},\ }\href {\doibase
  10.1021/cr9804691} {\bibfield  {journal} {\bibinfo  {journal} {Chemical
  Reviews}\ }\textbf {\bibinfo {volume} {100}},\ \bibinfo {pages} {1253}
  (\bibinfo {year} {2000})},\ \bibinfo {note} {pMID: 11749266},\ \Eprint
  {http://arxiv.org/abs/https://doi.org/10.1021/cr9804691}
  {https://doi.org/10.1021/cr9804691} \BibitemShut {NoStop}%
\bibitem [{\citenamefont {Okuda}(2023)}]{Okuda_2023}%
  \BibitemOpen
  \bibfield  {author} {\bibinfo {author} {\bibfnamefont {J.}~\bibnamefont
  {Okuda}},\ }\href {\doibase https://doi.org/10.1016/j.jorganchem.2023.122833}
  {\bibfield  {journal} {\bibinfo  {journal} {Journal of Organometallic
  Chemistry}\ }\textbf {\bibinfo {volume} {1000}},\ \bibinfo {pages} {122833}
  (\bibinfo {year} {2023})}\BibitemShut {NoStop}%
\bibitem [{\citenamefont {Jezequel}\ \emph {et~al.}(2001)\citenamefont
  {Jezequel}, \citenamefont {Dufaud}, \citenamefont {Ruiz-Garcia},
  \citenamefont {Carrillo-Hermosilla}, \citenamefont {Neugebauer},
  \citenamefont {Niccolai}, \citenamefont {Lefebvre}, \citenamefont {Bayard},
  \citenamefont {Corker}, \citenamefont {Fiddy}, \citenamefont {Evans},
  \citenamefont {Broyer}, \citenamefont {Malinge},\ and\ \citenamefont
  {Basset}}]{Jezequel_2001}%
  \BibitemOpen
  \bibfield  {author} {\bibinfo {author} {\bibfnamefont {M.}~\bibnamefont
  {Jezequel}}, \bibinfo {author} {\bibfnamefont {V.}~\bibnamefont {Dufaud}},
  \bibinfo {author} {\bibfnamefont {M.~J.}\ \bibnamefont {Ruiz-Garcia}},
  \bibinfo {author} {\bibfnamefont {F.}~\bibnamefont {Carrillo-Hermosilla}},
  \bibinfo {author} {\bibfnamefont {U.}~\bibnamefont {Neugebauer}}, \bibinfo
  {author} {\bibfnamefont {G.~P.}\ \bibnamefont {Niccolai}}, \bibinfo {author}
  {\bibfnamefont {F.}~\bibnamefont {Lefebvre}}, \bibinfo {author}
  {\bibfnamefont {F.}~\bibnamefont {Bayard}}, \bibinfo {author} {\bibfnamefont
  {J.}~\bibnamefont {Corker}}, \bibinfo {author} {\bibfnamefont
  {S.}~\bibnamefont {Fiddy}}, \bibinfo {author} {\bibfnamefont
  {J.}~\bibnamefont {Evans}}, \bibinfo {author} {\bibfnamefont {J.-P.}\
  \bibnamefont {Broyer}}, \bibinfo {author} {\bibfnamefont {J.}~\bibnamefont
  {Malinge}}, \ and\ \bibinfo {author} {\bibfnamefont {J.-M.}\ \bibnamefont
  {Basset}},\ }\href {\doibase 10.1021/ja000682q} {\bibfield  {journal}
  {\bibinfo  {journal} {Journal of the American Chemical Society}\ }\textbf
  {\bibinfo {volume} {123}},\ \bibinfo {pages} {3520} (\bibinfo {year}
  {2001})},\ \bibinfo {note} {pMID: 11472124},\ \Eprint
  {http://arxiv.org/abs/https://doi.org/10.1021/ja000682q}
  {https://doi.org/10.1021/ja000682q} \BibitemShut {NoStop}%
\bibitem [{\citenamefont {Ormaza}\ \emph
  {et~al.}(2017{\natexlab{a}})\citenamefont {Ormaza}, \citenamefont {Abufager},
  \citenamefont {Verlhac}, \citenamefont {Bachellier}, \citenamefont {Bocquet},
  \citenamefont {Lorente},\ and\ \citenamefont {Limot}}]{Ormaza_2017b}%
  \BibitemOpen
  \bibfield  {author} {\bibinfo {author} {\bibfnamefont {M.}~\bibnamefont
  {Ormaza}}, \bibinfo {author} {\bibfnamefont {P.}~\bibnamefont {Abufager}},
  \bibinfo {author} {\bibfnamefont {B.}~\bibnamefont {Verlhac}}, \bibinfo
  {author} {\bibfnamefont {N.}~\bibnamefont {Bachellier}}, \bibinfo {author}
  {\bibfnamefont {M.-L.}\ \bibnamefont {Bocquet}}, \bibinfo {author}
  {\bibfnamefont {N.}~\bibnamefont {Lorente}}, \ and\ \bibinfo {author}
  {\bibfnamefont {L.}~\bibnamefont {Limot}},\ }\href {\doibase
  10.1038/s41467-017-02151-6} {\bibfield  {journal} {\bibinfo  {journal}
  {Nature communications}\ } (\bibinfo {year} {2017}{\natexlab{a}}),\
  10.1038/s41467-017-02151-6}\BibitemShut {NoStop}%
\bibitem [{\citenamefont {Wei}\ \emph {et~al.}(2011)\citenamefont {Wei},
  \citenamefont {Sun}, \citenamefont {Benassi}, \citenamefont {Shen},
  \citenamefont {Sanvito},\ and\ \citenamefont {Hou}}]{Wei_2011}%
  \BibitemOpen
  \bibfield  {author} {\bibinfo {author} {\bibfnamefont {P.}~\bibnamefont
  {Wei}}, \bibinfo {author} {\bibfnamefont {L.}~\bibnamefont {Sun}}, \bibinfo
  {author} {\bibfnamefont {E.}~\bibnamefont {Benassi}}, \bibinfo {author}
  {\bibfnamefont {Z.}~\bibnamefont {Shen}}, \bibinfo {author} {\bibfnamefont
  {S.}~\bibnamefont {Sanvito}}, \ and\ \bibinfo {author} {\bibfnamefont
  {S.}~\bibnamefont {Hou}},\ }\href {\doibase 10.1063/1.3603446} {\bibfield
  {journal} {\bibinfo  {journal} {The Journal of Chemical Physics}\ }\textbf
  {\bibinfo {volume} {134}},\ \bibinfo {pages} {244704} (\bibinfo {year}
  {2011})},\ \Eprint
  {http://arxiv.org/abs/https://pubs.aip.org/aip/jcp/article-pdf/doi/10.1063/1.3603446/15440795/244704\_1\_online.pdf}
  {https://pubs.aip.org/aip/jcp/article-pdf/doi/10.1063/1.3603446/15440795/244704\_1\_online.pdf}
  \BibitemShut {NoStop}%
\bibitem [{\citenamefont {Zhang}\ \emph {et~al.}(2018)\citenamefont {Zhang},
  \citenamefont {Mu}, \citenamefont {Wei}, \citenamefont {Wang}, \citenamefont
  {Huang}, \citenamefont {Hu}, \citenamefont {Li},\ and\ \citenamefont
  {Wang}}]{Zhang_2018}%
  \BibitemOpen
  \bibfield  {author} {\bibinfo {author} {\bibfnamefont {G.-P.}\ \bibnamefont
  {Zhang}}, \bibinfo {author} {\bibfnamefont {Y.}~\bibnamefont {Mu}}, \bibinfo
  {author} {\bibfnamefont {M.-Z.}\ \bibnamefont {Wei}}, \bibinfo {author}
  {\bibfnamefont {S.}~\bibnamefont {Wang}}, \bibinfo {author} {\bibfnamefont
  {H.}~\bibnamefont {Huang}}, \bibinfo {author} {\bibfnamefont {G.-c.}\
  \bibnamefont {Hu}}, \bibinfo {author} {\bibfnamefont {Z.-L.}\ \bibnamefont
  {Li}}, \ and\ \bibinfo {author} {\bibfnamefont {C.}~\bibnamefont {Wang}},\
  }\href {\doibase 10.1039/C7TC05518H} {\bibfield  {journal} {\bibinfo
  {journal} {Journal of Materials Chemistry C}\ }\textbf {\bibinfo {volume}
  {6}},\ \bibinfo {pages} {2105} (\bibinfo {year} {2018})}\BibitemShut
  {NoStop}%
\bibitem [{\citenamefont {Ormaza}\ \emph
  {et~al.}(2017{\natexlab{b}})\citenamefont {Ormaza}, \citenamefont
  {Bachellier}, \citenamefont {Faraggi}, \citenamefont {Verlhac}, \citenamefont
  {Abufager}, \citenamefont {Ohresser}, \citenamefont {Joly}, \citenamefont
  {Romeo}, \citenamefont {Scheurer}, \citenamefont {Bocquet}, \citenamefont
  {Lorente},\ and\ \citenamefont {Limot}}]{Ormaza_2017a}%
  \BibitemOpen
  \bibfield  {author} {\bibinfo {author} {\bibfnamefont {M.}~\bibnamefont
  {Ormaza}}, \bibinfo {author} {\bibfnamefont {N.}~\bibnamefont {Bachellier}},
  \bibinfo {author} {\bibfnamefont {M.~N.}\ \bibnamefont {Faraggi}}, \bibinfo
  {author} {\bibfnamefont {B.}~\bibnamefont {Verlhac}}, \bibinfo {author}
  {\bibfnamefont {P.}~\bibnamefont {Abufager}}, \bibinfo {author}
  {\bibfnamefont {P.}~\bibnamefont {Ohresser}}, \bibinfo {author}
  {\bibfnamefont {L.}~\bibnamefont {Joly}}, \bibinfo {author} {\bibfnamefont
  {M.}~\bibnamefont {Romeo}}, \bibinfo {author} {\bibfnamefont
  {F.}~\bibnamefont {Scheurer}}, \bibinfo {author} {\bibfnamefont {M.-L.}\
  \bibnamefont {Bocquet}}, \bibinfo {author} {\bibfnamefont {N.}~\bibnamefont
  {Lorente}}, \ and\ \bibinfo {author} {\bibfnamefont {L.}~\bibnamefont
  {Limot}},\ }\href {https://api.semanticscholar.org/CorpusID:206738462}
  {\bibfield  {journal} {\bibinfo  {journal} {Nano letters}\ }\textbf {\bibinfo
  {volume} {17 3}},\ \bibinfo {pages} {1877} (\bibinfo {year}
  {2017}{\natexlab{b}})}\BibitemShut {NoStop}%
\bibitem [{\citenamefont {Czap}\ \emph {et~al.}(2019)\citenamefont {Czap},
  \citenamefont {Wagner}, \citenamefont {Xue}, \citenamefont {Gu},
  \citenamefont {Li}, \citenamefont {Yao}, \citenamefont {Wu},\ and\
  \citenamefont {Ho}}]{czap_probing_2019}%
  \BibitemOpen
  \bibfield  {author} {\bibinfo {author} {\bibfnamefont {G.}~\bibnamefont
  {Czap}}, \bibinfo {author} {\bibfnamefont {P.~J.}\ \bibnamefont {Wagner}},
  \bibinfo {author} {\bibfnamefont {F.}~\bibnamefont {Xue}}, \bibinfo {author}
  {\bibfnamefont {L.}~\bibnamefont {Gu}}, \bibinfo {author} {\bibfnamefont
  {J.}~\bibnamefont {Li}}, \bibinfo {author} {\bibfnamefont {J.}~\bibnamefont
  {Yao}}, \bibinfo {author} {\bibfnamefont {R.}~\bibnamefont {Wu}}, \ and\
  \bibinfo {author} {\bibfnamefont {W.}~\bibnamefont {Ho}},\ }\href {\doibase
  10.1126/science.aaw7505} {\bibfield  {journal} {\bibinfo  {journal}
  {Science}\ }\textbf {\bibinfo {volume} {364}},\ \bibinfo {pages} {670}
  (\bibinfo {year} {2019})}\BibitemShut {NoStop}%
\bibitem [{\citenamefont {Verlhac}\ \emph {et~al.}(2019)\citenamefont
  {Verlhac}, \citenamefont {Bachellier}, \citenamefont {Garnier}, \citenamefont
  {Ormaza}, \citenamefont {Abufager}, \citenamefont {Robles}, \citenamefont
  {Bocquet}, \citenamefont {Ternes}, \citenamefont {Lorente},\ and\
  \citenamefont {Limot}}]{Verlhac2019}%
  \BibitemOpen
  \bibfield  {author} {\bibinfo {author} {\bibfnamefont {B.}~\bibnamefont
  {Verlhac}}, \bibinfo {author} {\bibfnamefont {N.}~\bibnamefont {Bachellier}},
  \bibinfo {author} {\bibfnamefont {L.}~\bibnamefont {Garnier}}, \bibinfo
  {author} {\bibfnamefont {M.}~\bibnamefont {Ormaza}}, \bibinfo {author}
  {\bibfnamefont {P.}~\bibnamefont {Abufager}}, \bibinfo {author}
  {\bibfnamefont {R.}~\bibnamefont {Robles}}, \bibinfo {author} {\bibfnamefont
  {M.-L.}\ \bibnamefont {Bocquet}}, \bibinfo {author} {\bibfnamefont
  {M.}~\bibnamefont {Ternes}}, \bibinfo {author} {\bibfnamefont
  {N.}~\bibnamefont {Lorente}}, \ and\ \bibinfo {author} {\bibfnamefont
  {L.}~\bibnamefont {Limot}},\ }\href {\doibase 10.1126/science.aax8222}
  {\bibfield  {journal} {\bibinfo  {journal} {Science}\ }\textbf {\bibinfo
  {volume} {366}},\ \bibinfo {pages} {623} (\bibinfo {year}
  {2019})}\BibitemShut {NoStop}%
\bibitem [{\citenamefont {Braun}\ \emph {et~al.}(2006)\citenamefont {Braun},
  \citenamefont {Iancu}, \citenamefont {Pertaya}, \citenamefont {Rieder},\ and\
  \citenamefont {Hla}}]{Braun_2006}%
  \BibitemOpen
  \bibfield  {author} {\bibinfo {author} {\bibfnamefont {K.-F.}\ \bibnamefont
  {Braun}}, \bibinfo {author} {\bibfnamefont {V.}~\bibnamefont {Iancu}},
  \bibinfo {author} {\bibfnamefont {N.}~\bibnamefont {Pertaya}}, \bibinfo
  {author} {\bibfnamefont {K.-H.}\ \bibnamefont {Rieder}}, \ and\ \bibinfo
  {author} {\bibfnamefont {S.-W.}\ \bibnamefont {Hla}},\ }\href {\doibase
  10.1103/PhysRevLett.96.246102} {\bibfield  {journal} {\bibinfo  {journal}
  {Phys. Rev. Lett.}\ }\textbf {\bibinfo {volume} {96}},\ \bibinfo {pages}
  {246102} (\bibinfo {year} {2006})}\BibitemShut {NoStop}%
\bibitem [{\citenamefont {Ormaza}\ \emph {et~al.}(2015)\citenamefont {Ormaza},
  \citenamefont {Abufager}, \citenamefont {Bachellier}, \citenamefont {Robles},
  \citenamefont {Verot}, \citenamefont {Le~Bahers}, \citenamefont {Bocquet},
  \citenamefont {Lorente},\ and\ \citenamefont {Limot}}]{Ormaza_2015}%
  \BibitemOpen
  \bibfield  {author} {\bibinfo {author} {\bibfnamefont {M.}~\bibnamefont
  {Ormaza}}, \bibinfo {author} {\bibfnamefont {P.}~\bibnamefont {Abufager}},
  \bibinfo {author} {\bibfnamefont {N.}~\bibnamefont {Bachellier}}, \bibinfo
  {author} {\bibfnamefont {R.}~\bibnamefont {Robles}}, \bibinfo {author}
  {\bibfnamefont {M.}~\bibnamefont {Verot}}, \bibinfo {author} {\bibfnamefont
  {T.}~\bibnamefont {Le~Bahers}}, \bibinfo {author} {\bibfnamefont {M.-L.}\
  \bibnamefont {Bocquet}}, \bibinfo {author} {\bibfnamefont {N.}~\bibnamefont
  {Lorente}}, \ and\ \bibinfo {author} {\bibfnamefont {L.}~\bibnamefont
  {Limot}},\ }\href {\doibase 10.1021/jz5026118} {\bibfield  {journal}
  {\bibinfo  {journal} {The Journal of Physical Chemistry Letters}\ }\textbf
  {\bibinfo {volume} {6}},\ \bibinfo {pages} {395} (\bibinfo {year} {2015})},\
  \bibinfo {note} {pMID: 26261954},\ \Eprint
  {http://arxiv.org/abs/https://doi.org/10.1021/jz5026118}
  {https://doi.org/10.1021/jz5026118} \BibitemShut {NoStop}%
\bibitem [{\citenamefont {Ormaza}\ \emph {et~al.}(2016)\citenamefont {Ormaza},
  \citenamefont {Robles}, \citenamefont {Bachellier}, \citenamefont {Abufager},
  \citenamefont {Lorente},\ and\ \citenamefont {Limot}}]{Ormaza_2016}%
  \BibitemOpen
  \bibfield  {author} {\bibinfo {author} {\bibfnamefont {M.}~\bibnamefont
  {Ormaza}}, \bibinfo {author} {\bibfnamefont {R.}~\bibnamefont {Robles}},
  \bibinfo {author} {\bibfnamefont {N.}~\bibnamefont {Bachellier}}, \bibinfo
  {author} {\bibfnamefont {P.}~\bibnamefont {Abufager}}, \bibinfo {author}
  {\bibfnamefont {N.}~\bibnamefont {Lorente}}, \ and\ \bibinfo {author}
  {\bibfnamefont {L.}~\bibnamefont {Limot}},\ }\href {\doibase
  10.1021/acs.nanolett.5b04280} {\bibfield  {journal} {\bibinfo  {journal}
  {Nano Lett.}\ }\textbf {\bibinfo {volume} {16}},\ \bibinfo {pages} {588}
  (\bibinfo {year} {2016})}\BibitemShut {NoStop}%
\bibitem [{\citenamefont {Li}\ \emph {et~al.}(1992)\citenamefont {Li},
  \citenamefont {Hamrick}, \citenamefont {Van~Zee},\ and\ \citenamefont
  {Weltner}}]{Li_1992}%
  \BibitemOpen
  \bibfield  {author} {\bibinfo {author} {\bibfnamefont {S.}~\bibnamefont
  {Li}}, \bibinfo {author} {\bibfnamefont {Y.~M.}\ \bibnamefont {Hamrick}},
  \bibinfo {author} {\bibfnamefont {R.~J.}\ \bibnamefont {Van~Zee}}, \ and\
  \bibinfo {author} {\bibfnamefont {W.~J.}\ \bibnamefont {Weltner}},\ }\href
  {\doibase 10.1021/ja00037a078} {\bibfield  {journal} {\bibinfo  {journal}
  {Journal of the American Chemical Society}\ }\textbf {\bibinfo {volume}
  {114}},\ \bibinfo {pages} {4433} (\bibinfo {year} {1992})}\BibitemShut
  {NoStop}%
\bibitem [{\citenamefont {Bachellier}\ \emph {et~al.}(2016)\citenamefont
  {Bachellier}, \citenamefont {Ormaza}, \citenamefont {Faraggi}, \citenamefont
  {Verlhac}, \citenamefont {V\'erot}, \citenamefont {Le~Bahers}, \citenamefont
  {Bocquet},\ and\ \citenamefont {Limot}}]{Bachellier_2016}%
  \BibitemOpen
  \bibfield  {author} {\bibinfo {author} {\bibfnamefont {N.}~\bibnamefont
  {Bachellier}}, \bibinfo {author} {\bibfnamefont {M.}~\bibnamefont {Ormaza}},
  \bibinfo {author} {\bibfnamefont {M.}~\bibnamefont {Faraggi}}, \bibinfo
  {author} {\bibfnamefont {B.}~\bibnamefont {Verlhac}}, \bibinfo {author}
  {\bibfnamefont {M.}~\bibnamefont {V\'erot}}, \bibinfo {author} {\bibfnamefont
  {T.}~\bibnamefont {Le~Bahers}}, \bibinfo {author} {\bibfnamefont {M.-L.}\
  \bibnamefont {Bocquet}}, \ and\ \bibinfo {author} {\bibfnamefont
  {L.}~\bibnamefont {Limot}},\ }\href {\doibase 10.1103/PhysRevB.93.195403}
  {\bibfield  {journal} {\bibinfo  {journal} {Phys. Rev. B}\ }\textbf {\bibinfo
  {volume} {93}},\ \bibinfo {pages} {195403} (\bibinfo {year}
  {2016})}\BibitemShut {NoStop}%
\bibitem [{\citenamefont {Mier}\ \emph {et~al.}(2021)\citenamefont {Mier},
  \citenamefont {Verlhac}, \citenamefont {Garnier}, \citenamefont {Robles},
  \citenamefont {Limot}, \citenamefont {Lorente},\ and\ \citenamefont
  {Choi}}]{Mier_JPCL}%
  \BibitemOpen
  \bibfield  {author} {\bibinfo {author} {\bibfnamefont {C.}~\bibnamefont
  {Mier}}, \bibinfo {author} {\bibfnamefont {B.}~\bibnamefont {Verlhac}},
  \bibinfo {author} {\bibfnamefont {L.}~\bibnamefont {Garnier}}, \bibinfo
  {author} {\bibfnamefont {R.}~\bibnamefont {Robles}}, \bibinfo {author}
  {\bibfnamefont {L.}~\bibnamefont {Limot}}, \bibinfo {author} {\bibfnamefont
  {N.}~\bibnamefont {Lorente}}, \ and\ \bibinfo {author} {\bibfnamefont
  {D.-J.}\ \bibnamefont {Choi}},\ }\href {\doibase 10.1021/acs.jpclett.1c00328}
  {\bibfield  {journal} {\bibinfo  {journal} {The Journal of Physical Chemistry
  Letters}\ }\textbf {\bibinfo {volume} {12}},\ \bibinfo {pages} {2983}
  (\bibinfo {year} {2021})}\BibitemShut {NoStop}%
\bibitem [{\citenamefont {Welipitiya}\ \emph {et~al.}(1998)\citenamefont
  {Welipitiya}, \citenamefont {Waldfried}, \citenamefont {Borca}, \citenamefont
  {Dowben}, \citenamefont {Boag}, \citenamefont {Jiang}, \citenamefont
  {Gobulukoglu},\ and\ \citenamefont {Robertson}}]{Welipitiya_1998}%
  \BibitemOpen
  \bibfield  {author} {\bibinfo {author} {\bibfnamefont {D.}~\bibnamefont
  {Welipitiya}}, \bibinfo {author} {\bibfnamefont {C.}~\bibnamefont
  {Waldfried}}, \bibinfo {author} {\bibfnamefont {C.}~\bibnamefont {Borca}},
  \bibinfo {author} {\bibfnamefont {P.}~\bibnamefont {Dowben}}, \bibinfo
  {author} {\bibfnamefont {N.}~\bibnamefont {Boag}}, \bibinfo {author}
  {\bibfnamefont {H.}~\bibnamefont {Jiang}}, \bibinfo {author} {\bibfnamefont
  {I.}~\bibnamefont {Gobulukoglu}}, \ and\ \bibinfo {author} {\bibfnamefont
  {B.}~\bibnamefont {Robertson}},\ }\href {\doibase
  https://doi.org/10.1016/S0039-6028(98)00742-0} {\bibfield  {journal}
  {\bibinfo  {journal} {Surface Science}\ }\textbf {\bibinfo {volume} {418}},\
  \bibinfo {pages} {466} (\bibinfo {year} {1998})}\BibitemShut {NoStop}%
\bibitem [{\citenamefont {Pugmire}\ \emph {et~al.}(1999)\citenamefont
  {Pugmire}, \citenamefont {Woodbridge}, \citenamefont {Root},\ and\
  \citenamefont {Langell}}]{Pugmire_1999}%
  \BibitemOpen
  \bibfield  {author} {\bibinfo {author} {\bibfnamefont {D.~L.}\ \bibnamefont
  {Pugmire}}, \bibinfo {author} {\bibfnamefont {C.~M.}\ \bibnamefont
  {Woodbridge}}, \bibinfo {author} {\bibfnamefont {S.}~\bibnamefont {Root}}, \
  and\ \bibinfo {author} {\bibfnamefont {M.~A.}\ \bibnamefont {Langell}},\
  }\href {\doibase 10.1116/1.581854} {\bibfield  {journal} {\bibinfo  {journal}
  {Journal of Vacuum Science \& Technology A}\ }\textbf {\bibinfo {volume}
  {17}},\ \bibinfo {pages} {1581} (\bibinfo {year} {1999})},\ \Eprint
  {http://arxiv.org/abs/https://pubs.aip.org/avs/jva/article-pdf/17/4/1581/11020234/1581\_1\_online.pdf}
  {https://pubs.aip.org/avs/jva/article-pdf/17/4/1581/11020234/1581\_1\_online.pdf}
  \BibitemShut {NoStop}%
\bibitem [{\citenamefont {Pugmire}\ \emph {et~al.}(2001)\citenamefont
  {Pugmire}, \citenamefont {Woodbridge}, \citenamefont {Boag},\ and\
  \citenamefont {Langell}}]{Pugmire_2001}%
  \BibitemOpen
  \bibfield  {author} {\bibinfo {author} {\bibfnamefont {D.}~\bibnamefont
  {Pugmire}}, \bibinfo {author} {\bibfnamefont {C.}~\bibnamefont {Woodbridge}},
  \bibinfo {author} {\bibfnamefont {N.}~\bibnamefont {Boag}}, \ and\ \bibinfo
  {author} {\bibfnamefont {M.}~\bibnamefont {Langell}},\ }\href {\doibase
  https://doi.org/10.1016/S0039-6028(00)00939-0} {\bibfield  {journal}
  {\bibinfo  {journal} {Surface Science}\ }\textbf {\bibinfo {volume} {472}},\
  \bibinfo {pages} {155} (\bibinfo {year} {2001})}\BibitemShut {NoStop}%
\bibitem [{\citenamefont {Alessio}\ \emph {et~al.}(2023)\citenamefont
  {Alessio}, \citenamefont {Kotaru}, \citenamefont {Giudetti},\ and\
  \citenamefont {Krylov}}]{Alessio_2023}%
  \BibitemOpen
  \bibfield  {author} {\bibinfo {author} {\bibfnamefont {M.}~\bibnamefont
  {Alessio}}, \bibinfo {author} {\bibfnamefont {S.}~\bibnamefont {Kotaru}},
  \bibinfo {author} {\bibfnamefont {G.}~\bibnamefont {Giudetti}}, \ and\
  \bibinfo {author} {\bibfnamefont {A.~I.}\ \bibnamefont {Krylov}},\ }\href
  {\doibase 10.1021/acs.jpcc.2c05940} {\bibfield  {journal} {\bibinfo
  {journal} {The Journal of Physical Chemistry C}\ }\textbf {\bibinfo {volume}
  {127}},\ \bibinfo {pages} {3647} (\bibinfo {year} {2023})},\ \Eprint
  {http://arxiv.org/abs/https://doi.org/10.1021/acs.jpcc.2c05940}
  {https://doi.org/10.1021/acs.jpcc.2c05940} \BibitemShut {NoStop}%
\bibitem [{\citenamefont {Edmondson}\ and\ \citenamefont
  {Saywell}(2022)}]{Edmondson_2022}%
  \BibitemOpen
  \bibfield  {author} {\bibinfo {author} {\bibfnamefont {M.}~\bibnamefont
  {Edmondson}}\ and\ \bibinfo {author} {\bibfnamefont {A.}~\bibnamefont
  {Saywell}},\ }\href {\doibase 10.1021/acs.nanolett.2c02895} {\bibfield
  {journal} {\bibinfo  {journal} {Nano Letters}\ }\textbf {\bibinfo {volume}
  {22}},\ \bibinfo {pages} {8210} (\bibinfo {year} {2022})},\ \bibinfo {note}
  {publisher: American Chemical Society}\BibitemShut {NoStop}%
\bibitem [{\citenamefont {Meyer}\ \emph {et~al.}(1996)\citenamefont {Meyer},
  \citenamefont {Baikie}, \citenamefont {Kopatzki},\ and\ \citenamefont
  {Behm}}]{Meyer_1996}%
  \BibitemOpen
  \bibfield  {author} {\bibinfo {author} {\bibfnamefont {J.~A.}\ \bibnamefont
  {Meyer}}, \bibinfo {author} {\bibfnamefont {I.~D.}\ \bibnamefont {Baikie}},
  \bibinfo {author} {\bibfnamefont {E.}~\bibnamefont {Kopatzki}}, \ and\
  \bibinfo {author} {\bibfnamefont {R.~J.}\ \bibnamefont {Behm}},\ }\href
  {\doibase 10.1016/0039-6028(96)00852-7} {\bibfield  {journal} {\bibinfo
  {journal} {Surface Science}\ }\textbf {\bibinfo {volume} {365}},\ \bibinfo
  {pages} {L647} (\bibinfo {year} {1996})}\BibitemShut {NoStop}%
\bibitem [{\citenamefont {Repp}\ \emph {et~al.}(2000)\citenamefont {Repp},
  \citenamefont {Moresco}, \citenamefont {Meyer}, \citenamefont {Rieder},
  \citenamefont {Hyldgaard},\ and\ \citenamefont {Persson}}]{Repp_2000}%
  \BibitemOpen
  \bibfield  {author} {\bibinfo {author} {\bibfnamefont {J.}~\bibnamefont
  {Repp}}, \bibinfo {author} {\bibfnamefont {F.}~\bibnamefont {Moresco}},
  \bibinfo {author} {\bibfnamefont {G.}~\bibnamefont {Meyer}}, \bibinfo
  {author} {\bibfnamefont {K.-H.}\ \bibnamefont {Rieder}}, \bibinfo {author}
  {\bibfnamefont {P.}~\bibnamefont {Hyldgaard}}, \ and\ \bibinfo {author}
  {\bibfnamefont {M.}~\bibnamefont {Persson}},\ }\href {\doibase
  10.1103/PhysRevLett.85.2981} {\bibfield  {journal} {\bibinfo  {journal}
  {Phys. Rev. Lett.}\ }\textbf {\bibinfo {volume} {85}},\ \bibinfo {pages}
  {2981} (\bibinfo {year} {2000})}\BibitemShut {NoStop}%
\bibitem [{\citenamefont {Sotthewes}\ \emph {et~al.}(2021)\citenamefont
  {Sotthewes}, \citenamefont {Nijmeijer},\ and\ \citenamefont
  {Zandvliet}}]{Sotthewes}%
  \BibitemOpen
  \bibfield  {author} {\bibinfo {author} {\bibfnamefont {K.}~\bibnamefont
  {Sotthewes}}, \bibinfo {author} {\bibfnamefont {M.}~\bibnamefont
  {Nijmeijer}}, \ and\ \bibinfo {author} {\bibfnamefont {H.~J.~W.}\
  \bibnamefont {Zandvliet}},\ }\href {\doibase 10.1103/PhysRevB.103.245311}
  {\bibfield  {journal} {\bibinfo  {journal} {Phys. Rev. B}\ }\textbf {\bibinfo
  {volume} {103}},\ \bibinfo {pages} {245311} (\bibinfo {year}
  {2021})}\BibitemShut {NoStop}%
\bibitem [{\citenamefont {Fernandez-Torrente}\ \emph
  {et~al.}(2007)\citenamefont {Fernandez-Torrente}, \citenamefont {Monturet},
  \citenamefont {Franke}, \citenamefont {Fraxedas}, \citenamefont {Lorente},\
  and\ \citenamefont {Pascual}}]{Fernandez-Torrente_2007}%
  \BibitemOpen
  \bibfield  {author} {\bibinfo {author} {\bibfnamefont {I.}~\bibnamefont
  {Fernandez-Torrente}}, \bibinfo {author} {\bibfnamefont {S.}~\bibnamefont
  {Monturet}}, \bibinfo {author} {\bibfnamefont {K.~J.}\ \bibnamefont
  {Franke}}, \bibinfo {author} {\bibfnamefont {J.}~\bibnamefont {Fraxedas}},
  \bibinfo {author} {\bibfnamefont {N.}~\bibnamefont {Lorente}}, \ and\
  \bibinfo {author} {\bibfnamefont {J.~I.}\ \bibnamefont {Pascual}},\ }\href
  {\doibase 10.1103/PhysRevLett.99.176103} {\bibfield  {journal} {\bibinfo
  {journal} {Physical Review Letters}\ }\textbf {\bibinfo {volume} {99}},\
  \bibinfo {pages} {176103} (\bibinfo {year} {2007})}\BibitemShut {NoStop}%
\bibitem [{\citenamefont {Mielke}\ \emph {et~al.}(2015)\citenamefont {Mielke},
  \citenamefont {Hanke}, \citenamefont {Peters}, \citenamefont {Hecht},
  \citenamefont {Persson},\ and\ \citenamefont {Grill}}]{Mielke_2014}%
  \BibitemOpen
  \bibfield  {author} {\bibinfo {author} {\bibfnamefont {J.}~\bibnamefont
  {Mielke}}, \bibinfo {author} {\bibfnamefont {F.}~\bibnamefont {Hanke}},
  \bibinfo {author} {\bibfnamefont {M.~V.}\ \bibnamefont {Peters}}, \bibinfo
  {author} {\bibfnamefont {S.}~\bibnamefont {Hecht}}, \bibinfo {author}
  {\bibfnamefont {M.}~\bibnamefont {Persson}}, \ and\ \bibinfo {author}
  {\bibfnamefont {L.}~\bibnamefont {Grill}},\ }\href {\doibase
  10.1021/ja510528x} {\bibfield  {journal} {\bibinfo  {journal} {Journal of the
  American Chemical Society}\ }\textbf {\bibinfo {volume} {137}},\ \bibinfo
  {pages} {1844} (\bibinfo {year} {2015})},\ \bibinfo {note} {pMID: 25494667},\
  \Eprint {http://arxiv.org/abs/https://doi.org/10.1021/ja510528x}
  {https://doi.org/10.1021/ja510528x} \BibitemShut {NoStop}%
\bibitem [{\citenamefont {Robles}\ \emph {et~al.}(2020)\citenamefont {Robles},
  \citenamefont {Zobač}, \citenamefont {Yeung}, \citenamefont {Moresco},
  \citenamefont {Joachim},\ and\ \citenamefont {Lorente}}]{Robles_2020}%
  \BibitemOpen
  \bibfield  {author} {\bibinfo {author} {\bibfnamefont {R.}~\bibnamefont
  {Robles}}, \bibinfo {author} {\bibfnamefont {V.}~\bibnamefont {Zobač}},
  \bibinfo {author} {\bibfnamefont {K.~H.~A.}\ \bibnamefont {Yeung}}, \bibinfo
  {author} {\bibfnamefont {F.}~\bibnamefont {Moresco}}, \bibinfo {author}
  {\bibfnamefont {C.}~\bibnamefont {Joachim}}, \ and\ \bibinfo {author}
  {\bibfnamefont {N.}~\bibnamefont {Lorente}},\ }\href {\doibase
  10.1039/D0CP01657H} {\bibfield  {journal} {\bibinfo  {journal} {Physical
  Chemistry Chemical Physics}\ }\textbf {\bibinfo {volume} {22}},\ \bibinfo
  {pages} {15208} (\bibinfo {year} {2020})},\ \bibinfo {note} {publisher: The
  Royal Society of Chemistry}\BibitemShut {NoStop}%
\bibitem [{\citenamefont {Berger}\ \emph {et~al.}(2022)\citenamefont {Berger},
  \citenamefont {Jeindl}, \citenamefont {H\"ormann},\ and\ \citenamefont
  {Hofmann}}]{Berger_2022}%
  \BibitemOpen
  \bibfield  {author} {\bibinfo {author} {\bibfnamefont {R.~K.}\ \bibnamefont
  {Berger}}, \bibinfo {author} {\bibfnamefont {A.}~\bibnamefont {Jeindl}},
  \bibinfo {author} {\bibfnamefont {L.}~\bibnamefont {H\"ormann}}, \ and\
  \bibinfo {author} {\bibfnamefont {O.~T.}\ \bibnamefont {Hofmann}},\ }\href
  {\doibase 10.1021/acs.jpcc.2c00994} {\bibfield  {journal} {\bibinfo
  {journal} {The Journal of Physical Chemistry C}\ }\textbf {\bibinfo {volume}
  {126}},\ \bibinfo {pages} {7718} (\bibinfo {year} {2022})},\ \Eprint
  {http://arxiv.org/abs/https://doi.org/10.1021/acs.jpcc.2c00994}
  {https://doi.org/10.1021/acs.jpcc.2c00994} \BibitemShut {NoStop}%
\bibitem [{\citenamefont {Kresse}\ and\ \citenamefont
  {Furthm\"uller}(1996)}]{kresse_efficiency_1996}%
  \BibitemOpen
  \bibfield  {author} {\bibinfo {author} {\bibfnamefont {G.}~\bibnamefont
  {Kresse}}\ and\ \bibinfo {author} {\bibfnamefont {J.}~\bibnamefont
  {Furthm\"uller}},\ }\href {\doibase 10.1016/0927-0256(96)00008-0} {\bibfield
  {journal} {\bibinfo  {journal} {Comput. Mater. Sci.}\ }\textbf {\bibinfo
  {volume} {6}},\ \bibinfo {pages} {15} (\bibinfo {year} {1996})}\BibitemShut
  {NoStop}%
\bibitem [{\citenamefont {Kresse}\ and\ \citenamefont
  {Joubert}(1999)}]{kresse_ultrasoft_1999}%
  \BibitemOpen
  \bibfield  {author} {\bibinfo {author} {\bibfnamefont {G.}~\bibnamefont
  {Kresse}}\ and\ \bibinfo {author} {\bibfnamefont {D.}~\bibnamefont
  {Joubert}},\ }\href {\doibase 10.1103/PhysRevB.59.1758} {\bibfield  {journal}
  {\bibinfo  {journal} {Phys. Rev. B}\ }\textbf {\bibinfo {volume} {59}},\
  \bibinfo {pages} {1758} (\bibinfo {year} {1999})}\BibitemShut {NoStop}%
\bibitem [{\citenamefont {Perdew}\ \emph {et~al.}(1996)\citenamefont {Perdew},
  \citenamefont {Burke},\ and\ \citenamefont
  {Ernzerhof}}]{perdew_generalized_1996}%
  \BibitemOpen
  \bibfield  {author} {\bibinfo {author} {\bibfnamefont {J.~P.}\ \bibnamefont
  {Perdew}}, \bibinfo {author} {\bibfnamefont {K.}~\bibnamefont {Burke}}, \
  and\ \bibinfo {author} {\bibfnamefont {M.}~\bibnamefont {Ernzerhof}},\ }\href
  {\doibase 10.1103/PhysRevLett.77.3865} {\bibfield  {journal} {\bibinfo
  {journal} {Phys. Rev. Lett.}\ }\textbf {\bibinfo {volume} {77}},\ \bibinfo
  {pages} {3865} (\bibinfo {year} {1996})}\BibitemShut {NoStop}%
\bibitem [{\citenamefont {Tkatchenko}\ and\ \citenamefont
  {Scheffler}(2009)}]{tkatchenko_accurate_2009}%
  \BibitemOpen
  \bibfield  {author} {\bibinfo {author} {\bibfnamefont {A.}~\bibnamefont
  {Tkatchenko}}\ and\ \bibinfo {author} {\bibfnamefont {M.}~\bibnamefont
  {Scheffler}},\ }\href {\doibase 10.1103/PhysRevLett.102.073005} {\bibfield
  {journal} {\bibinfo  {journal} {Phys. Rev. Lett.}\ }\textbf {\bibinfo
  {volume} {102}},\ \bibinfo {pages} {073005} (\bibinfo {year}
  {2009})}\BibitemShut {NoStop}%
\bibitem [{\citenamefont {Tersoff}\ and\ \citenamefont
  {Hamann}(1985)}]{tersoff_theory_1985}%
  \BibitemOpen
  \bibfield  {author} {\bibinfo {author} {\bibfnamefont {J.}~\bibnamefont
  {Tersoff}}\ and\ \bibinfo {author} {\bibfnamefont {D.~R.}\ \bibnamefont
  {Hamann}},\ }\href {\doibase 10.1103/PhysRevB.31.805} {\bibfield  {journal}
  {\bibinfo  {journal} {Phys. Rev. B}\ }\textbf {\bibinfo {volume} {31}},\
  \bibinfo {pages} {805} (\bibinfo {year} {1985})}\BibitemShut {NoStop}%
\bibitem [{\citenamefont {Bocquet}\ \emph {et~al.}(2009)\citenamefont
  {Bocquet}, \citenamefont {Lesnard}, \citenamefont {Monturet},\ and\
  \citenamefont {Lorente}}]{bocquet_theory_2009}%
  \BibitemOpen
  \bibfield  {author} {\bibinfo {author} {\bibfnamefont {M.-L.}\ \bibnamefont
  {Bocquet}}, \bibinfo {author} {\bibfnamefont {H.}~\bibnamefont {Lesnard}},
  \bibinfo {author} {\bibfnamefont {S.}~\bibnamefont {Monturet}}, \ and\
  \bibinfo {author} {\bibfnamefont {N.}~\bibnamefont {Lorente}},\ }in\
  \href@noop {} {\emph {\bibinfo {booktitle} {Computational {Methods} in
  {Catalysis} and {Materials} {Science}}}},\ \bibinfo {editor} {edited by\
  \bibinfo {editor} {\bibfnamefont {R.~A.~v.}\ \bibnamefont {Santen}}\ and\
  \bibinfo {editor} {\bibfnamefont {P.}~\bibnamefont {Sautet}}}\ (\bibinfo
  {publisher} {Wiley-VCH Verlag GmbH \& Co. KGaA},\ \bibinfo {year} {2009})\
  pp.\ \bibinfo {pages} {199--219}\BibitemShut {NoStop}%
\bibitem [{\citenamefont {Lorente}\ and\ \citenamefont
  {Robles}(2019)}]{lorente_stmpw_2019}%
  \BibitemOpen
  \bibfield  {author} {\bibinfo {author} {\bibfnamefont {N.}~\bibnamefont
  {Lorente}}\ and\ \bibinfo {author} {\bibfnamefont {R.}~\bibnamefont
  {Robles}},\ }\href {\doibase 10.5281/zenodo.3581159} {\enquote {\bibinfo
  {title} {{STMpw v1.0b2 (Zenodo)}},}\ }\bibinfo {howpublished}
  {https://doi.org/10.5281/zenodo.3581159} (\bibinfo {year} {2019})\BibitemShut
  {NoStop}%
\bibitem [{\citenamefont {Momma}\ and\ \citenamefont
  {Izumi}(2011)}]{momma_vesta_2011}%
  \BibitemOpen
  \bibfield  {author} {\bibinfo {author} {\bibfnamefont {K.}~\bibnamefont
  {Momma}}\ and\ \bibinfo {author} {\bibfnamefont {F.}~\bibnamefont {Izumi}},\
  }\href {\doibase 10.1107/S0021889811038970} {\bibfield  {journal} {\bibinfo
  {journal} {Journal of Applied Crystallography}\ }\textbf {\bibinfo {volume}
  {44}},\ \bibinfo {pages} {1272} (\bibinfo {year} {2011})}\BibitemShut
  {NoStop}%
\end{thebibliography}%

\end{document}